\documentclass[journal]{IEEEtran}

\usepackage{mathtools}
\usepackage{bm}
\usepackage{threeparttable}
\usepackage{hyperref}
\usepackage{flushend}

\usepackage{array}
\usepackage{booktabs}

\newtheorem{proposition}{Proposition}
\usepackage{nicefrac}
\usepackage{enumitem}
\usepackage{algorithm}
\usepackage{algorithmicx}
\usepackage{algpseudocode}
\usepackage[input-ignore={,},input-decimal-markers={.}]{siunitx}
\usepackage[dvipsnames]{xcolor}

\usepackage{tikz}
\usepackage{circuitikz}
\usepackage{pgfplots}
\pgfplotsset{compat=1.14}

\usetikzlibrary{calc}
\usetikzlibrary{shapes}
\usetikzlibrary{shapes.misc}
\usetikzlibrary{matrix}
\usetikzlibrary{arrows}
\usetikzlibrary{decorations.markings}
\usetikzlibrary{spy}
\usetikzlibrary{decorations.pathmorphing}
\usetikzlibrary{decorations.markings}
\usetikzlibrary{positioning}
\usetikzlibrary{fit}
\usetikzlibrary{petri}
\usetikzlibrary{chains}
\usetikzlibrary{intersections}
\usetikzlibrary{plotmarks}
\usetikzlibrary{pgfplots.groupplots}
\usetikzlibrary{backgrounds}
\usetikzlibrary{pgfplots.colormaps}

\newcommand{\fixme}[2]{\ifx&#2&{\color{red}#1}\else{\color{red}FIXME\{}#1{\color{red}\}}\footnote{{\color{red}#2}}\PackageWarning{Fixme}{#1: #2}\fi}

\definecolor{Set1-7-1}{RGB}{228,26,28}
\definecolor{Set1-7-2}{RGB}{55,126,184}
\definecolor{Set1-7-3}{RGB}{77,175,74}
\definecolor{Set1-7-4}{RGB}{152,78,163}
\definecolor{Set1-7-5}{RGB}{255,127,0}
\definecolor{Set1-7-6}{RGB}{166,86,40}
\definecolor{Set1-7-7}{RGB}{0,0,0}

\usepackage{pgfplots}
\usepgfplotslibrary{units}

\pgfplotsset{
    legend image with text/.style={
        legend image code/.code={%
            \node[anchor=center] at (0.3cm,0cm) {#1};
        }
    },
}

\pgfplotsset{
/pgfplots/colormap={mymap1}{
rgb255=(8,29,88) 
rgb255=(37,52,148) 
rgb255=(34,94,168)
rgb255=(29,145,192) 
rgb255=(65,182,196) 
rgb255=(127,205,187) 
rgb255=(199,233,180)
rgb255=(237,248,177) 
rgb255=(255,255,217)}}

\pgfplotsset{
/pgfplots/colormap={mymap2}{
rgb255=(1,70,54) 
rgb255=(1,108,89) 
rgb255=(2,129,138)
rgb255=(54,144,192) 
rgb255=(103,169,207) 
rgb255=(166,189,219) 
rgb255=(208,209,230)
rgb255=(236,226,240) 
rgb255=(255,247,251)}}

\pgfplotsset{
/pgfplots/colormap={mymap3}{
rgb255=(128,0,38) 
rgb255=(189,0,38) 
rgb255=(227,26,28)
rgb255=(252,78,42) 
rgb255=(253,141,60) 
rgb255=(254,178,76) 
rgb255=(254,217,118)
rgb255=(255,237,160) 
rgb255=(255,255,204)}}

\pgfplotsset{
/pgfplots/colormap={mymap4}{
rgb255=(166,206,227) 
rgb255=(31,120,180)
rgb255=(178,223,138) 
rgb255=(51,160,44)}}

\IEEEoverridecommandlockouts

\usepackage{todonotes}
\presetkeys{todonotes}{inline}{}
\usepackage{mathrsfs}
\usepackage{amssymb}
\usepackage{gensymb}
\makeatletter
\let\MYcaption\@makecaption
\makeatother
\usepackage{subcaption}
\makeatletter
\let\@makecaption\MYcaption
\makeatother

\begin{document}

% Remove dashes in repeated names
\bstctlcite{IEEEexample:BSTcontrol}

\title{Hardware Implementation of Neural Self-Interference Cancellation}

\author{\IEEEauthorblockN{Yann Kurzo, %
Andreas Toftegaard Kristensen, %
Andreas Burg,~\IEEEmembership{Member,~IEEE,}\\ %
Alexios Balatsoukas-Stimming,~\IEEEmembership{Member,~IEEE}%
}%
\thanks{Y. Kurzo is with ON Semiconductor, 2074 Marin-Epagnier, Switzerland (e-mail: yann.kurzo@gmail.com).}%
\thanks{A. Kristensen and A. Burg are with the Telecommunications Circuits Laboratory, \'Ecole polytechnique f\'ed\'erale de Lausanne, 1015 Lausanne, Switzerland (e-mail: \{andreas.kristensen,andreas.burg\}@epfl.ch).}%
\thanks{A. Balatsoukas-Stimming is with the Eindhoven University of Technology, 5600 MB Eindhoven, The Netherlands (e-mail: a.k.balatsoukas.stimming@tue.nl).}
\thanks{Parts of this work were presented at the 2018 Asilomar Conference on Signals, Systems, and Computers~\cite{Kurzo2018}.}%
\thanks{This work was supported by the Swiss National Science Foundation under project \#200021\_182621.}
}

\maketitle

\begin{abstract}
In-band full-duplex systems can transmit and receive information simultaneously and on the same frequency band. However, due to the strong self-interference caused by the transmitter to its own receiver, the use of non-linear digital self-interference cancellation is essential. In this work, we describe a hardware architecture for a neural network-based non-linear self-interference (SI) canceller and we compare it with our own hardware implementation of a conventional polynomial based SI canceller. 
Our results show that, for the same SI cancellation performance, the neural network canceller has an $8.1\times$ smaller area and requires $7.7\times$ less power than the polynomial canceller. Moreover, the neural network canceller can achieve $7$~dB more SI cancellation while still being $1.2\times$ smaller than the polynomial canceller and only requiring $1.3\times$ more power. These results show that NN-based methods applied to communications are not only useful from a performance perspective, but can also lead to order-of-magnitude implementation complexity reductions.
\end{abstract}


\section{Introduction}
In-band full-duplex (FD) communications have for long been considered to be impractical due to the strong self-interference (SI) caused by the transmitter to its own receiver. However, recent work on the topic (e.g.,~\cite{Jain2011,Duarte2012,Bharadia2013}) has demonstrated that it is, in fact, possible to achieve sufficient SI cancellation (SIC) to make FD systems viable. Typically, SIC is performed in both the radio frequency (RF) domain and the digital domain to cancel the SI signal down to the level of the receiver noise floor. There are several RF cancellation methods, that can be broadly categorized into \emph{passive RF cancellation} and \emph{active RF cancellation}. Some form of RF cancellation is generally necessary to avoid saturating the analog front-end of the receiver. Passive RF cancellation can be obtained by using, e.g., circulators, directional antennas, beamforming, polarization, or shielding~\cite{Everett2014}. Active RF cancellation is commonly implemented by transforming the transmitted RF signal appropriately to emulate the SI channel using analog components and subtracting the resulting SIC signal from the received SI signal~\cite{Jain2011,Bharadia2013}. Alternatively, an additional transmitter can be used to generate the SIC signal from the transmitted baseband samples~\cite{Duarte2012}.

However, a residual SI signal is typically still present at the receiver after RF cancellation has been performed. This residual SI signal can, in principle, be easily canceled in the digital domain, since it is caused by a known transmitted signal. Unfortunately, in practice, several transceiver non-linearities distort the SI signal. Some examples of non-linearities include baseband non-linearities (e.g., digital-to-analog converter (DAC) and analog-to-digital converter (ADC))~\cite{Balatsoukas2015}, IQ imbalance~\cite{Balatsoukas2015,Korpi2014}, phase-noise~\cite{Sahai2013,Syrjala2014}, and power amplifier (PA) non-linearities~\cite{Balatsoukas2015,Korpi2014,Anttila2014,Korpi2017}. These effects need to be taken into account using intricate polynomial models to cancel the SI to the level of the receiver noise floor. These polynomial models perform well in practice, but their implementation complexity grows rapidly with the maximum considered non-linearity order. Principal component analysis (PCA) is an effective complexity reduction technique that can identify the most significant non-linearity terms in a parallel Hammerstein model~\cite{Korpi2017}. However, with PCA-based methods, the transmitted digital baseband samples need to be multiplied with a transformation matrix to generate the SIC signal, thus introducing additional complexity. Moreover, whenever the SI channel changes, the high-complexity PCA operation needs to be re-run. To the best of our knowledge, no hardware implementation of a polynomial SI canceller has been reported in the open literature to date. Only the work of~\cite{Campo2018} has made a step in this direction, since the authors considered quantization aspects of polynomial SI cancellers.

In the past few years, there has been renewed interest in the use of neural networks (NNs) to augment or replace a range of signal processing tasks in communications systems~\cite{OShea2017a,Wang2017a,Mao2018a,Gunduz2019,Qin2019a,Balatsoukas2019sips}. NNs are particularly well-suited to tackle non-linear signal processing problems, where traditional model-based algorithms are unavailable or too complex for analytical treatment. However, NN-based solutions can also be used in cases where traditional model-based algorithms suffer from prohibitively high implementation complexity. For example, NNs have been used to successfully perform digital predistortion (DPD) in wireless systems~\cite{Tarver2019b,Hongyo2019}, non-linear leakage cancellation in FDD transceivers~\cite{Ploder2019}, as well as optical fiber non-linearity compensation~\cite{Hager2018a}. NNs have also been used for non-linear SIC in full-duplex communications~\cite{Balatsoukas2018,Guo2018,Kristensen2019} and it was shown in~\cite{Balatsoukas2018} that they can achieve similar SIC performance with a state-of-the-art polynomial SIC model, but with much lower complexity.

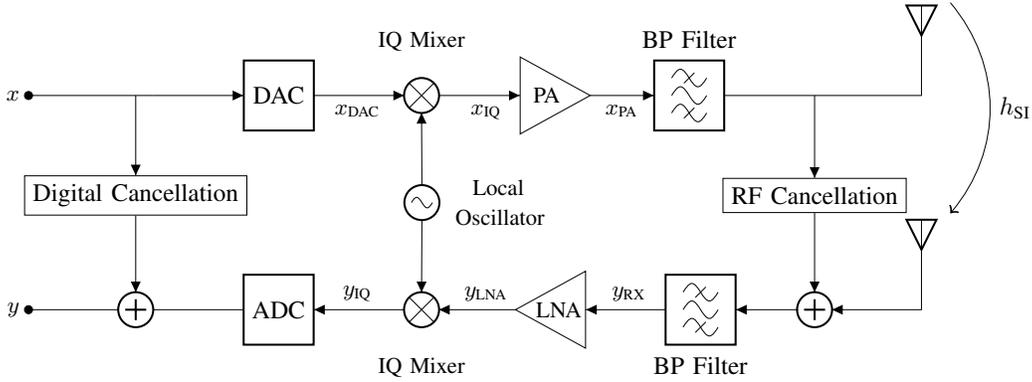
\begin{figure*}[t]
  \centering
  \scalebox{0.95}{\begin{circuitikz}[scale=1]

	%%% Transmitter %%%	
	% Nodes
	\draw (4,0) node[mixer,scale=0.5] (txmixer) {};
	\draw (11,0) node[antenna,scale=0.6] (txantenna) {};
	\draw (0,-1.4) node[draw] (digcanc) {Digital Cancellation};
	
	% Connections	
	\draw (-1.5,0) to[short,*-] (0.25,0);
	\draw (0.25,0) to[short,-] ++(0,0) to[twoport,>,t=DAC] (txmixer.west) node[inputarrow]{};
	\draw (txmixer.east) to[short,-] ++(0.5,0) to[amp,>,t=\small{PA}] ++(2.25,0) to[bandpass,>,l^=BP Filter] ++(1.5,0) to ++(1.5,0) to (txantenna);
	
	% Text labels
	\draw (-1.5,0) node[left] {\small $x$};
	\draw (txmixer)+(0,0.5) node[above] {\small{IQ Mixer}};
	\draw (txmixer)+(-0.9,0) node[below] {\small $x_{\text{DAC}}$};
	\draw (txmixer)+(0.9,0) node[below] {\small $x_{\text{IQ}}$};
	\draw (txantenna)+(-4.2,0) node[below] {\small $x_{\text{PA}}$};
	
	%%% Receiver %%%
	% Nodes
	\draw (11,-3) node[antenna,scale=0.6] (rxantenna) {};
	\draw (9.5,-3) node[adder,scale=0.5] (rxadder) {};
	\draw (4,-3) node[mixer,scale=0.5] (rxmixer) {};
	\draw (txantenna)++(-1.5,-1.4) node[draw] (rfcanc) {RF Cancellation};
	\draw (0,-3) node[adder,scale=0.5] (rxadderdig) {};
	
	% Connections	
	% Digital canceler
	\draw (0,0) to (digcanc.north) node[inputarrow,rotate=-90]{};
	\draw (digcanc.south) to (rxadderdig.north) node[inputarrow,rotate=-90]{};
	% RF canceler
	\draw (txantenna)++(-1.5,0) to (rfcanc.north) node[inputarrow,rotate=-90]{};
	\draw (rfcanc.south) to (rxadder.north) node[inputarrow,rotate=-90]{};
	\draw (rxantenna) to[short] (rxadder.east) node[inputarrow,rotate=180]{};
	% Rest of chain
	\draw (rxadder.west) to ++(0,0) to[bandpass,>,l=BP Filter] ++(-2.7,0) to[amp,>,t={\rotatebox[origin=c]{180}{\small{LNA}}}] ++(-1.5,0) to[short,-] (rxmixer.east) node[inputarrow,rotate=180]{};
	\draw (rxmixer.west) to[twoport,>,t=ADC] (rxadderdig.east);
	\draw (rxadderdig.west) to[short,-*] (-1.5,-3);
	\draw [->] (txantenna)+(0.4,1.3) to[thick, out=-50, in=50, edge node={node [right] {$h_{\text{SI}}$}}]  ($(rxantenna) + (0.4,1.35)$);
	
	% Text labels
	\draw (rxantenna)+(-4.1,0) node[above] {\small $y_{\text{RX}}$};
	\draw (rxmixer)+(0.9,0) node[above] {\small $y_{\text{LNA}}$};
	\draw (rxmixer)+(-0.9,0) node[above] {\small $y_{\text{IQ}}$};
	\draw (-1.5,-3) node[left] {\small $y$};
	\draw (rxmixer)+(0,-0.55) node[below] {\small{IQ Mixer}};
	
	%%% Reference clock %%%
	\draw (4.25,-1.5) node[oscillator,scale=0.5] (ref) {};
	\draw (ref.south) to[short,-] (rxmixer.north) node[inputarrow,rotate=270]{};
	\draw (ref.north) to[short,-] (txmixer.south) node[inputarrow,rotate=90]{};
	\draw (ref)+(+0.1,0) node[right,text width=1.2cm,align=center] {\small{Local\\Oscillator}};
	
\end{circuitikz}}
  \caption{Block diagram of a full-duplex transceiver with active RF SIC and digital SIC. A few components have been omitted for simplicity, a more detailed diagram can be found in~\cite{Korpi2017}.}\label{fig:block}
\end{figure*}

The communications subsystems of embedded devices are typically implemented using dedicated hardware instead of general-purpose processors. The main reason for this is that using dedicated hardware enables very high energy efficiency, which is particularly desirable in mobile platforms. As such, dedicated hardware solutions for NN-assisted communications systems are essential. Existing NN hardware accelerators, such as~\cite{Zhang2015,Chen2016}, mainly target applications where both the size of the NN and the number of inputs is very large, and where producing a few tens of outputs per second is sufficient. Communications applications, on the other hand, use relatively small NNs with few inputs, but need to provide millions of outputs per second. As such, communications applications generally require more highly parallelized NN hardware accelerator architectures. However, to date, only a small number of works have considered these hardware-related issues in the context of communications applications. Specifically, the works of~\cite{Aoudia2019,Wodiany2019} study NN quantization as a first step towards hardware implementation, while the authors of \cite{Kurzo2018,Tarver2019a} describe actual hardware implementations of simple NNs for SIC in full-duplex communications and DPD, respectively. 
%\looseness=-1

\subsubsection*{Contribution}
In this work, we present a hardware implementation of the SIC method proposed in~\cite{Balatsoukas2018} to quantify and translate the computational complexity gains over the state-of-the-art polynomial based model of~\cite{Korpi2017} into real-world hardware resource utilization gains. 
Contrary to~\cite{Kurzo2018,Balatsoukas2018}, we use a more realistic and well-defined measurement setup that results in a completely new and more challenging dataset with more non-linear content.
Since, to the best of our knowledge, no polynomial SI canceller implementations have been reported in the literature, we also present a hardware architecture for a reference polynomial SI canceller. We note that this hardware architecture can also be used for other related applications such as digital predistortion and leakage cancellation in FDD transceivers. We provide FPGA and ASIC implementation results that clearly demonstrate the significant gains with respect to the polynomial SI canceller that can be achieved by an NN-based SI canceller in terms of resource utilization, throughput, and energy efficiency.
\subsubsection*{Outline}
The remainder of this paper is organized as follows. Section~\ref{sec:background} provides background on full-duplex communications and digital SIC using polynomial cancellers, while Section~\ref{sec:nncanc} describes how SIC can be achieved using NNs. In Section~\ref{sec:nnhardware}, we describe our proposed NN-based SI canceller hardware architecture and, in Section~\ref{sec:polyhardware}, we describe our proposed baseline polynomial-based SI canceller hardware architecture. In Section~\ref{sec:sicresults}, we compare the performance and the complexity of a conventional polynomial SI canceller with the NN-based SI cancellers. In Section~\ref{sec:hardwareresults}, we also provide FPGA and ASIC implementation results. Finally, Section~\ref{sec:conclusion} concludes this paper.
\section{Conventional Digital Self-Interference Cancellation}\label{sec:background}

Fig.~\ref{fig:block} shows a block diagram of a full-duplex transceiver. On the transmitter side, the digital baseband samples $x[n] \in \mathbb{C}$, where $n$ is the sample index, are converted to an analog signal using a DAC, up-converted to a carrier frequency $f_c$ using an IQ mixer, amplified using a power amplifier (PA), and filtered using a bandpass (BP) filter. The transmitted signal leaks to the receiver through an SI channel $h_{\text{SI}}$ and is then filtered using a BP filter, amplified using an LNA, downconverted using an IQ mixer, and digitized using an ADC. The SI channel, $h_{\text{SI}}$, also models the passive RF SIC. An RF cancellation signal is subtracted from the received SI signal at some point before the LNA to avoid saturating the receiver. Since the transmitter and the receiver are co-located, they share a common local oscillator (LO) signal to minimize the effect of phase noise on the SI signal~\cite{Syrjala2014}.

If we assume, for simplicity of exposition, that there is no signal-of-interest from a remote node and no thermal noise, then the received signal $y[n]$ in Fig.~\ref{fig:block} consists only of the residual SI signal after RF SIC has been performed. We denote the received signal in this special case by $y_{\text{SI}}[n]$. The goal of digital SIC is to reproduce an accurate copy of $y_{\text{SI}}[n]$, denoted by $\hat{y}_{\text{SI}}[n]$, based on samples of the transmitted baseband signal $x[n]$. This signal is then subtracted from $y[n]$ so that the residual SI signal is $y_{\text{SI}}[n] - \hat{y}_{\text{SI}}[n]$. If $\hat{y}_{\text{SI}}[n]$ is reconstructed perfectly, then the SI can be canceled entirely and $y_{\text{SI}}[n] - \hat{y}_{\text{SI}}[n] = 0$. In practice, as discussed previously, due to the presence of thermal noise and transceiver non-linearities, perfect SIC is difficult to achieve. The SIC performance $C_{\text{dB}}$ is typically evaluated as:
\begin{align}
  C_{\text{dB}} &= 10 \log_{10} \left( \frac{\sum_n |y_{\text{SI}}[n]|^2}{\sum_n |y_{\text{SI}}[n] - \hat{y}_{\text{SI}}[n]|^2} \right).
\end{align}

\subsection{Linear Self-Interference Cancellation}\label{sec:lincanc}
Linear SIC is the simplest form of SIC that ignores all non-linear effects of the various components in Fig.~\ref{fig:block}. The linear SIC signal is constructed as~\cite{Bharadia2013}:
\begin{align}
	\hat{y}_{\text{SI}}[n] & = \sum _{l=0}^{L-1}\hat{h}[l]x[n-l], \label{eq:lincanc}
\end{align}
where $\hat{h}[l] \in \mathbb{C},~l \in \{0,\hdots,L-1\},$ models the SI channel, $h_{\text{SI}}$, and any other memory effect in the transceiver chain. The parameters $\hat{h}[l]$ can be obtained from training samples either in a one-shot fashion using standard least-squares (LS) estimation or adaptively using an iterative version of the LS estimation algorithm, such as least mean squares (LMS) or recursive least squares (RLS).

\subsection{Polynomial Non-Linear Self-Interference Cancellation}\label{sec:polycanc}
Each active component in the transceiver model shown in Fig.~\ref{fig:block} is generally a dynamic non-linear system. This means that linear cancellation alone is, in most cases, not accurate enough to cancel a sufficiently large fraction of the SI signal. It has been shown that the transmitter IQ imbalance and the PA non-linearities typically dominate all remaining non-linearities~\cite{Anttila2014,Korpi2017}. This is true in particular when the transmitter and receiver chains use the same local oscillator signal for upconversion, as shown in Fig.~\ref{fig:block}, so that the effect of phase noise becomes negligible~\cite{Syrjala2014}. As such, the SIC signal $\hat{y}_{\text{SI}}[n]$ can be constructed as~\cite{Anttila2014,Korpi2017}:
\begin{align}
	\hat{y}_{\text{SI}}[n]	& = \sum _{\substack{p=1,\\p \text{ odd}}}^P \sum_{q=0}^p\sum_{l=0}^{L-1}\hat{h}_{p,q}[l] x[n-l]^{q}x^*[n-l]^{p-q}, \label{eq:final}
\end{align}
where $\hat{h}_{p,q}[l] \in \mathbb{C}$ and only odd values for $p$ are considered because even harmonics typically lie out-of-band and are filtered out by the transmitter and receiver BP filters. 
%capture the joint effect of $\hat{h}_{p}[l]$ and of the IQ imbalance parameters $K_1$ and $K_2$. 
The model in \eqref{eq:final} is linear with respect to the parameters $\hat{h}_{p,q}[l]$, and therefore, similarly to linear SI estimation, the parameters $\hat{h}_{p,q}[l]$ can be estimated based on training samples using some variant of the LS estimation algorithm. The \emph{basis functions} of the polynomial model in \eqref{eq:final} are defined as:
\begin{align}
  \text{BF}_{p,q}(x) & = x^{q}(x^*)^{p-q}. \label{eq:basisfunction}
\end{align}
The number of distinct basis functions in~\eqref{eq:final} is~\cite{Korpi2017}:
\begin{align}
	N_{\text{BF}}	& = \frac{L}{4}\left(P+1\right)\left(P+3\right). \label{eq:nbf}
\end{align}
Using \eqref{eq:basisfunction}, the expression for $\hat{y}[n]$ in \eqref{eq:final} can be re-written in a more compact form:
\begin{align}
	\hat{y}_{\text{SI}}[n]	& = \sum _{\substack{p=1,\\p \text{ odd}}}^P \sum_{q=0}^p\sum_{l=0}^{L-1}\hat{h}_{p,q}[l] \text{BF}_{p,q}\left(x[n-l]\right). \label{eq:finalbf}
\end{align}
We note that linear cancellation is a special case of the polynomial model in~\eqref{eq:finalbf} when only considering the single term for $p=1$ and $q=1$.

\subsection{Computational Complexity}
A multiplication between two complex numbers $x_1 = a +jb$ and $x_2 = c+jd$ can be performed in a straightforward manner as $x_1x_2 = (ac-bd) + j(ad+bc)$. This requires two real-valued additions and four real-valued multiplications. However, it is also possible to first compute $s_1 = ac$, $s_2 = bd$, and $s_3 = (a+b)(c+d)$ so that $x_1x_2 = (s_1-s_2) + j(s_3-s_1-s_2)$. This requires five real-valued additions and three real-valued multiplications. Since hardware multipliers are typically significantly more complex than hardware adders, we assume the latter method is used to minimize the number of multipliers.
Thus, it can directly be deduced from \eqref{eq:lincanc} that the total number of real-valued multiplications and additions that are required by the linear SI canceller is:
\begin{align}
	N_{\text{ADD,lin}}	& = 7L - 2,\\
	N_{\text{MUL,lin}}	& = 3L.
\end{align}
Moreover, if we ignore the computation of the basis functions for simplicity,\footnote{We note that this simplification is justified in Section~\ref{sec:polyhardware}.} the total number of real-valued multiplications and additions that are required by the polynomial SI canceller (which also includes the linear cancellation term) is~\cite{Balatsoukas2018}:
\begin{align}
	N_{\text{ADD,poly}}	& = \frac{7}{4}L\left(P+1\right)\left(P+3\right)-2, \label{eq:addpoly}\\
	N_{\text{MUL,poly}}	& = \frac{3}{4}L\left(P+1\right)\left(P+3\right). \label{eq:multpoly}
\end{align}
We note that the expression for $N_{\text{ADD,poly}}$ in our previous work of~\cite{Balatsoukas2018} erroneously ignored the five real-valued additions that are required to implement each complex multiplication. As such, the actual complexity of the polynomial canceller is even higher than that reported in~\cite{Balatsoukas2018}.

\section{Neural Network Non-Linear Digital Self-Interference Cancellation}\label{sec:nncanc}

Polynomial SIC models such as~\eqref{eq:finalbf} work well in practice but are often highly redundant in the sense that many of the $\hat{h}_{p,q}[l]$ parameters are very close to zero. NN-based SI cancellers, on the other hand, can extract the essence of the non-linear structure of the SI signal from training data, which often significantly reduces the complexity of the SIC model~\cite{Balatsoukas2018}. A challenge when using NN cancellers is that the NN training process is inherently noisy due to the use of mini-batches for gradient estimation, which makes it difficult to achieve a very accurate reconstruction of the SI signal~\cite{Kristensen2019}. To overcome this problem, \cite{Balatsoukas2018} used a NN to reconstruct only a particular part of the SI signal, while using conventional linear cancellation for the remainder of the SI. Specifically, in \cite{Balatsoukas2018} the SI signal was conceptually decomposed into a linear component and a non-linear component:
\begin{align}
	y_{\text{SI}}[n]	& = y_{\text{SI,linear}}[n] + y_{\text{SI,nl}}[n].
\end{align}
The SIC is carried out in two steps. First, linear cancellation is used to reconstruct $\hat{y}_{\text{SI,linear}}[n]$ as:
\begin{align}
  \hat{y}_{\text{SI,linear}}[n] & = \sum_{l=0}^{L-1}\hat{h}[l] x[n-l]. \label{eq:linear}
\end{align}
The parameters $\hat{h}[l]$ are obtained using LS estimation while considering the substantially weaker signal $\hat{y}_{\text{SI,nl}}[n]$ as noise. The linear SIC signal is then subtracted from the SI signal to obtain:
\begin{align}
	y_{\text{SI,nl}}[n] & \approx y_{\text{SI}}[n] - \hat{y}_{\text{SI,linear}}[n].
\end{align}
The task of the NN is limited to reconstructing $y_{\text{SI,nl}}[n]$ based on the appropriate $x[n]$ samples. 

As is common practice when training NNs, we normalize the input and output training samples so that $x[n]$ and $y_{\text{SI,nl}}[n]$ have unit variance (i.e., the variance of the real part and the variance of the imaginary part are both equal to $0.5$) and zero mean. To perform SIC on the test data, the output of the NN is denormalized using the mean and variance estimated based on the training data.

\subsection{Neural Network Structure}
\begin{figure}[t]
  \centering
  \scalebox{0.925}{\hspace{-0.75cm}\def\layersep{3.05cm}
\def\twolayersep{6.1cm}
\def\inputnodes{6}
\def\hiddennodes{5}
\def\outputnodes{2}

\begin{tikzpicture}[shorten >=1pt,->,draw=black!75, node distance=\layersep, scale=0.55]
    \tikzstyle{every pin edge}=[<-,shorten <=1pt]
    \tikzstyle{neuron}=[circle,draw=black!75,fill=black!75,minimum size=11pt,inner sep=0pt]
    \tikzstyle{input neuron}=[neuron, fill=black!20!green];
    \tikzstyle{output neuron}=[neuron, fill=red!75];
    \tikzstyle{hidden neuron}=[neuron, fill=blue!75];
    \tikzstyle{annot} = [text width=10em, text centered]

    % Draw the input layer nodes
    %\foreach \name / \y in {1,...,\inputnodes}
    % This is the same as writing \foreach \name / \y in {1/1,2/2,3/3,4/4}
	\node[input neuron, pin=left: {\small$\Re{\left\{x[n]\right\}}$}] (I-1) at (0,-1) {};
	\node[input neuron, pin=left: {\small$\Im{\left\{x[n]\right\}}$}] (I-2) at (0,-2) {};
	\node[input neuron, pin=left: {\small$\Re{\left\{x[n{-}1]\right\}}$}] (I-3) at (0,-3) {};
	\node[input neuron, pin=left: {\small$\Im{\left\{x[n{-}1]\right\}}$}] (I-4) at (0,-4) {};
	\node at (-2.5,-4.85) {$\vdots$};
	\node at (0,-4.75) {$\vdots$};
	\node[input neuron, pin=left: {\small$\Re{\left\{x[n{-}L+1]\right\}}$}] (I-5) at (0,-6) {};
	\node[input neuron, pin=left: {\small$\Im{\left\{x[n{-}L+1]\right\}}$}] (I-6) at (0,-7) {};

    % Draw the hidden layer nodes
	\pgfmathsetmacro{\limit}{\hiddennodes-1}
	\foreach \name / \y in {1,...,\limit}
		\path[yshift=-0.5cm]
			node[hidden neuron] (H-\name) at (\layersep,-\y cm) {};
	\node at (\layersep,-5.25) {$\vdots$};
	\path[yshift=-0.5cm]
		node[hidden neuron] (H-\hiddennodes) at (\layersep,-6 cm) {};

    % Draw the output layer node
	\foreach \name / \y in {1,...,\outputnodes}
		\path[yshift=-2cm]
			node[output neuron,pin={[pin edge={->}]right:\ifodd\y {\small $\Re{\left\{\hat{y}_{\text{SI,non-linear}}[n]\right\}}$} \else {\small $\Im{\left\{\hat{y}_{\text{SI,non-linear}}[n]\right\}}$} \fi}] (O-\name) at (\twolayersep,-\y) {};

    % Connect every node in the input layer with every node in the
    % hidden layer.
    \foreach \source in {1,...,\inputnodes}
        \foreach \dest in {1,...,\hiddennodes}
            %\path (I-\source) edge node [pos=0.75, sloped, above] {\scriptsize $w^{(1)}_{\dest,\source}$} (H-\dest);
			\path (I-\source) edge (H-\dest);

    % Connect every node in the hidden layer with the output layer
    \foreach \source in {1,...,\hiddennodes}
		\foreach \dest in {1,...,\outputnodes}
			%\path (H-\source) edge node [pos=0.25, sloped, above] {\scriptsize $w^{(2)}_{\dest,\source}$} (O-\dest);
			\path (H-\source) edge (O-\dest);

\end{tikzpicture}
% End of code
}
  \caption{Example of a neural network with one hidden layer for the reconstruction of the non-linear component $y_{\text{SI,nl}}[n]$ of the SI signal~\cite{Balatsoukas2018}.}\label{fig:nn}
\end{figure}
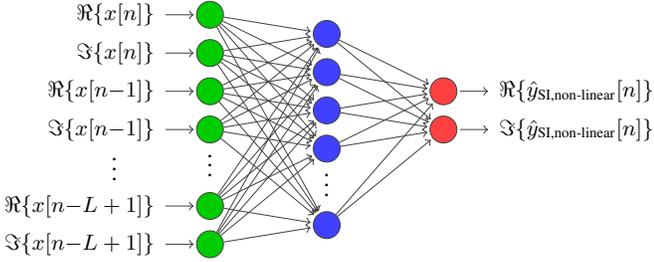

Due to the universal approximation theorem~\cite{Hornik1991}, a feed-forward NN with one hidden layer, as depicted in Fig.~\ref{fig:nn}, is sufficient to reconstruct the non-linear SI signal. While the work of~\cite{Balatsoukas2018} only considered feedforward NNs with one hidden layer, it is possible to use any NN architecture to generate $\hat{y}_{\text{SI,nl}}[n]$. In particular,~\cite{Kristensen2019} employed a deep feedforward NN and showed that using many layers with few neurons per layer can have significant computational complexity advantages with respect to a shallow NN SI canceller that uses a single layer with more neurons. In all cases and as shown in Fig.~\ref{fig:nn}, the cancellation NNs have $2L$ input nodes, which correspond to the real and imaginary parts of the $L$ delayed versions of $x[n]$, and two output nodes, which correspond to the real and imaginary parts of the target $\hat{y}_{\text{SI,nl}}[n]$ sample. In the following, we denote the number of hidden layers by $N_l$ and the number of hidden nodes per layer $N_h$.
Let the vector $\mathbf{l}_0$ contain the $2L$ inputs to the NN:
\begingroup
\setlength\arraycolsep{1.25pt}
\begin{align}
	\mathbf{l}_0 & = \begin{bmatrix}
		\Re{\left\{x[n]\right\}} & \Im{\left\{x[n]\right\}} & \hdots & \Re{\left\{x[n{-}L{+}1]\right\}} & \Im{\left\{x[n{-}L{+}1]\right\}}
	\end{bmatrix}^{T}.
\end{align}
\endgroup
The outputs of the first hidden layer neurons are given by:
\begin{align}
	\mathbf{l}_1	& = f_1\left(\mathbf{W}_{1}\mathbf{l}_{0} + \mathbf{b}_{1}\right), \label{eq:layer1}
\end{align}
where $\mathbf{W}_{1}$ is an $N_h \times 2L$ matrix containing the hidden layer weights, $\mathbf{b}_{1}$ is an $N_h \times 1$ vector containing the hidden layer biases, and $f_1(\cdot)$ is the (vectorized) non-linear activation function used in the first hidden layer. The outputs of the neurons in the hidden layers $1 < l \leq N_l$ are:
\begin{align}
	\mathbf{l}_l	& = f_l\left(\mathbf{W}_{l}\mathbf{l}_{l-1} + \mathbf{b}_{l}\right), \label{eq:layerl}
\end{align}
where $\mathbf{W}_{l}$ is an $N_h \times N_h$ matrix containing the hidden layer weights, $\mathbf{b}_{l}$ is an $N_h \times 1$ vector containing the hidden layer biases, and $f_l(\cdot)$ is the (vectorized) non-linear activation function used in hidden layer $l$. Finally, the outputs of the output layer neurons are given by:
\begin{align}
	\mathbf{l}_{N_l+1}	& = f_{N_l+1}\left(\mathbf{W}_{N_l+1}\mathbf{l}_{N_l} + \mathbf{b}_{N_l+1}\right), \label{eq:layero}
\end{align}
where $\mathbf{W}_{N_l+1}$ is a $2 \times N_h$ matrix containing the output layer weights, $\mathbf{b}_{N_l+1}$ is a $2 \times 1$ vector containing the output layer biases, and $f_{N_l+1}$ is the activation function used in the output layer. As can be seen in Fig.~\ref{fig:nn}, for $\mathbf{l}_{N_l+1}$ we have:
\begin{align}
	\mathbf{l}_{N_l+1} & = 
	\begin{bmatrix}
		\Re{\left\{\hat{y}_{\text{nl}}[n]\right\}} &
		\Im{\left\{\hat{y}_{\text{nl}}[n]\right\}}
	\end{bmatrix}^T.
\end{align}
The goal of the NN is to minimize the mean squared error between the expected NN output and the actual NN output:
\begin{align}
	\text{MSE}	& = \frac{1}{N}\sum_{n=0}^{N-1}\left(\Re{\left\{y_{\text{SI,nl}}[n]\right\}}-\Re{\left\{\hat{y}_{\text{SI,nl}}[n]\right\}}\right)^2 \nonumber \\
				& + \frac{1}{N}\sum_{n=0}^{N-1}\left(\Im{\left\{y_{\text{SI,nl}}[n]\right\}}-\Im{\left\{\hat{y}_{\text{SI,nl}}[n]\right\}}\right)^2, \label{eq:MSE}
\end{align}
where $N$ is the total number of training samples. The MSE in~\eqref{eq:MSE} is minimized by choosing appropriate values for $\mathbf{W}_{l},~\mathbf{b}_{l},~l \in \{1,\hdots,N_l+1\}$, using back-propagation~\cite{Rumelhart1986}.

\subsection{Computational Complexity}
Let us assume that the NN uses the popular ReLU activation function in the hidden layers (which has similar complexity to a real-valued addition (i.e., $f_l = \text{ReLU}(\mathbf{x}) = \max(\mathbf{0},\mathbf{x})$, $l \in \{1,\hdots,N_l\}$) and a linear activation function in the output layer (i.e., $f_{N_l+1}(\mathbf{x}) = \mathbf{x}$). Then, the number of real-valued multiplications and additions that are required by a NN canceller with a single hidden layer and $N_h$ hidden neurons is~\cite{Balatsoukas2018}:
\begin{align}
	N_{\text{ADD,NN}}	& = (2L+3)N_h + 7L,  \label{eq:addnnsingle}\\
	N_{\text{MUL,NN}}	& = (2L+2)N_h + 3L, \label{eq:multnnsingle}
\end{align}
where the second term in both expressions comes from the linear SI canceller that is required for the NN SI canceller to work. Moreover, two additions are required to add the output of the linear SI canceller with the output of the NN canceller.\footnote{We note that these two additions were not included in~\cite{Balatsoukas2018}, but we include them here for the sake of accuracy.} For the more general NN described in~\cite{Kristensen2019} with $N_l$ hidden layers with $N_h$ neurons each, \eqref{eq:addnnsingle}-\eqref{eq:multnnsingle} can be generalized to:
\begin{align}
	N_{\text{ADD,NN}}	& = (2L+3 + (N_l{-}1)(N_h{+}1))N_h + 7L, \label{eq:addnn}\\
	N_{\text{MUL,NN}}	& = (2L + 2 + (N_l{-}1)N_h)N_h + 3L. \label{eq:multnn}
\end{align}
\begin{figure}[t!]
	\centering
	\scalebox{0.6}{\def\horizontalsep{4.5cm}
\def\vertsep{3.5cm}

\def\horizontalsepsmall{1.5cm}
\def\vertsepsmall{1.5cm}

\def\filterheight{1.54cm}
\def\filterwidth{2.5cm}

\begin{tikzpicture}[-triangle 45, scale=0.65, thick]
  \tikzstyle{dspfilter}=[rectangle, draw=black, minimum width=\filterwidth, minimum height=\filterheight, inner sep=4pt, thick, align=center]
	
	%%%% NODES %%%%
	\node[dspfilter] (yn) at (0,0) {$y[n]$ \\ \\ \textbf{Real and} \\ \textbf{Imaginary}};
	\node[dspfilter] (xn) at ($(yn.south)+ (0,-\vertsep)$) {$x[n]$ \\ $x[n-1]$ \\ $\vdots$ \\ $x[n-L+1]$  \\ \\ \textbf{Real and} \\ \textbf{Imaginary}};

	\node[dspfilter] (linear) at ($(xn.east)+ (\horizontalsep,\vertsepsmall)$) {\textbf{Linear} \\ \textbf{Approximator} \\ $\hat{h}$};    	
	
	\node[dspfilter] (nn) at ($(xn.east)+ (\horizontalsep,-\vertsepsmall)$) {\textbf{Neural} \\ \textbf{Network} \\ $w_{i,j}$ and $b_j$};    	
	
	\node[dspfilter] (denorm) at ($(nn.east)+ (\horizontalsep,0)$) {\textbf{Denormalization}};  	

	\node[circle, draw=black,inner sep=0pt, thick] (sum1) at ($(linear.east)!0.5!(nn.east) + (\horizontalsep, \vertsepsmall)$) {\Large $\bm{+}$};	
		
	\node[circle, draw=black,inner sep=0pt, thick] (sum2) at (sum1 |- yn) {\Large $\bm{-}$};		
		
	\node[dspfilter] (output) at ($(sum2) + (\horizontalsep,0)$) {$y_c[n]$ \\ \textbf{Real and} \\ \textbf{Imaginary}};
	
	%%%% CONNECTIONS %%%%
  
	\draw[-triangle 45] (yn) -- node [above, pos=0.95] {$+$} (sum2);
	
	\draw[-triangle 45] (xn) -- ($(linear)!0.5!(nn) + (-0.75*\horizontalsep,0)$) |- (linear);
	\draw[-triangle 45] (xn) -- ($(linear)!0.5!(nn) + (-0.75*\horizontalsep,0)$) |- (nn);		
	\node[thick, circle, fill=black, minimum width=0.175cm, inner sep=0] at ($(linear)!0.5!(nn) + (-0.75*\horizontalsep,0)$) {};			
	
	\draw[-triangle 45] (nn) -- node [below, pos=0.5] {$\hat{y}_{\text{nn}}[n]$} (denorm);	
	
	\draw[-triangle 45] (linear) -- node [below, pos=0.5] {$\hat{y}_{\text{lin}}[n]$} (sum1);		
	
	\draw[-triangle 45] (denorm) -- (sum1);		
	
	\draw[-triangle 45] (sum1) -- node [right, pos=0.5] {$\hat{y}[n]$} (sum2);			
	
	\draw[-triangle 45] (sum2) -- (output);				
	
\end{tikzpicture}}
	\caption{High-level architecture on the NN-based SIC scheme~\cite{Kurzo2018}.}
	\label{fig:si-canceller-nn}
\end{figure}
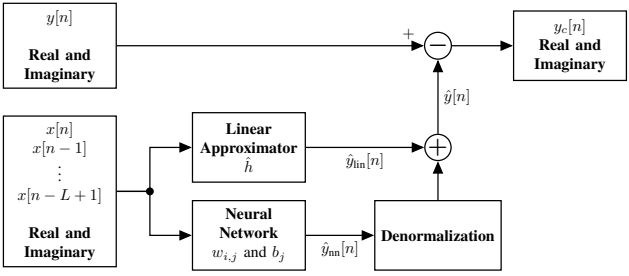
\section{Neural Network Canceller Hardware Architecture}\label{sec:nnhardware}
In this section, we describe a generic hardware architecture that can be used to implement both the shallow NN-based SI canceller of~\cite{Balatsoukas2018} and deeper NN-based SI cancellers such as the ones described in~\cite{Kristensen2019}. We first provide an overview of the architecture, which is followed by a more detailed explanation of each component. In Fig.~\ref{fig:si-canceller-nn}, we show the high-level architecture of a general NN-based canceller. The set of baseband samples $\left\{x[n],\hdots,x[n-L+1]\right\}$ is given as an input to a linear SI canceller and a NN-based SI canceller. These two SI cancellers operate in parallel to generate the linear and non-linear cancellation signals, respectively, which are then added (after the denormalization step for the NN) to produce the cancellation signal $\hat{y}_{\text{SI}}[n]$.
\subsection{Macro-Pipeline Architecture}
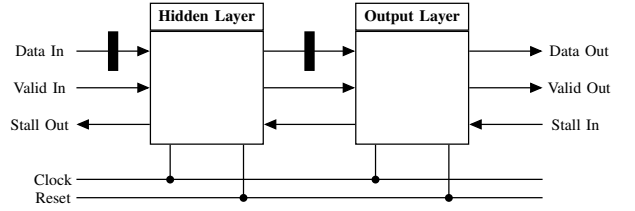
\begin{figure}[t!]
	\centering
	\scalebox{0.6}{\def\horizontalsep{7.0cm}
\def\vertsep{3.5cm}

\def\horizontalsepsmall{2.5cm}
\def\vertsepsmall{1.25cm}

\def\blockwidth{2.5cm}
\def\topblockheight{0.5cm}
\def\bottomblockheight{2.5cm}

\begin{tikzpicture}[-triangle 45, scale=0.65, thick]

  \tikzstyle{topblock}=[rectangle, draw=black, minimum width=\blockwidth, minimum height=\topblockheight, inner sep=4pt, thick, align=center]
  \tikzstyle{bottomblock}=[rectangle, draw=black, minimum width=\blockwidth, minimum height=\bottomblockheight, inner sep=4pt, thick, align=center]	
	
	%%%% NODES %%%%
	\node[topblock] (hidden_top) at (0,0) {\textbf{Hidden Layer}};
	\node[bottomblock, anchor = north] (hidden_bot) at (hidden_top.south) {};   

	\node[topblock] (output_top) at ($(hidden_top) + (\horizontalsep, 0)$) {\textbf{Output Layer}};
	\node[bottomblock, anchor = north] (output_bot) at (output_top.south) {};   

	%%%% CONNECTIONS %%%%
  
	\draw[-triangle 45] ($(hidden_bot.west) + (-\horizontalsepsmall, \vertsepsmall)$) -- node [pos=-0.5] {{Data In}} ($(hidden_bot.west) + (0, \vertsepsmall)$);
	\draw[-triangle 45] ($(hidden_bot.west) + (-\horizontalsepsmall, 0)$) -- node [pos=-0.5] {{Valid In}} ($(hidden_bot.west) + (0, 0)$);
	\draw[-triangle 45] ($(hidden_bot.west) + (0, -\vertsepsmall)$)  -- node [pos=1.5] {{Stall Out}}  ($(hidden_bot.west) + (-\horizontalsepsmall, -\vertsepsmall)$);

	\draw[-triangle 45] ($(hidden_bot.east) + (0, \vertsepsmall)$) -- ($(output_bot.west) + (0, \vertsepsmall)$);
	\draw[-triangle 45] ($(hidden_bot.east) + (0, 0)$) -- (output_bot.west);
	\draw[-triangle 45] ($(output_bot.west) + (0, -\vertsepsmall)$) -- ($(hidden_bot.east) + (0, -\vertsepsmall)$) ;

	\draw[-triangle 45] ($(output_bot.east) + (0, \vertsepsmall)$) -- node [pos=1.5] {{Data Out}}  ($(output_bot.east) + (\horizontalsepsmall, \vertsepsmall)$);
	\draw[-triangle 45] ($(output_bot.east) + (0, 0)$) -- node [pos=1.5] {{Valid Out}}  ($(output_bot.east) + (\horizontalsepsmall, 0)$);
	\draw[-triangle 45] ($(output_bot.east) + (\horizontalsepsmall, -\vertsepsmall)$) -- node [pos=-0.45] {{Stall In}}  ($(output_bot.east) + (0, -\vertsepsmall)$) ;

	\draw[-] ($(hidden_bot.west) + (-\horizontalsepsmall, -2.5*\vertsepsmall)$) -- node [pos=-0.05] {{Clock}}  ($(output_bot.east) + (\horizontalsepsmall, -2.5*\vertsepsmall)$);
	\draw[-] ($(hidden_bot.west) + (-\horizontalsepsmall, -3*\vertsepsmall)$) -- node [pos=-0.05] {{Reset}}  ($(output_bot.east) + (\horizontalsepsmall, -3*\vertsepsmall)$);

	% Clock Connections
	\draw[-] ($(hidden_bot.south) + (-0.5*\horizontalsepsmall, 0)$) -- ($(hidden_bot.south) + (-0.5*\horizontalsepsmall, -0.95*\vertsepsmall)$);
	\node[thick, circle, fill=black, minimum width=0.175cm, inner sep=0] at ($(hidden_bot.south) + (-0.5*\horizontalsepsmall, -0.95*\vertsepsmall)$) {};				
	
	\draw[-] ($(output_bot.south) + (-0.5*\horizontalsepsmall, 0)$) -- ($(output_bot.south) + (-0.5*\horizontalsepsmall, -0.95*\vertsepsmall)$);
	\node[thick, circle, fill=black, minimum width=0.175cm, inner sep=0] at ($(output_bot.south) + (-0.5*\horizontalsepsmall, -0.95*\vertsepsmall)$) {};				

	% Reset Connections
	\draw[-] ($(hidden_bot.south) + (0.5*\horizontalsepsmall, 0)$) -- ($(hidden_bot.south) + (0.5*\horizontalsepsmall, -1.45*\vertsepsmall)$);
	\node[thick, circle, fill=black, minimum width=0.175cm, inner sep=0] at ($(hidden_bot.south) + (0.5*\horizontalsepsmall, -1.45*\vertsepsmall)$) {};					
	
	\draw[-] ($(output_bot.south) + (0.5*\horizontalsepsmall, 0)$) -- ($(output_bot.south) + (0.5*\horizontalsepsmall, -1.45*\vertsepsmall)$);
	\node[thick, circle, fill=black, minimum width=0.175cm, inner sep=0] at ($(output_bot.south) + (0.5*\horizontalsepsmall, -1.45*\vertsepsmall)$) {};					
	
	% Squares for clock on the lines
	\node [draw, fill=black, thick, shape=rectangle, minimum width=0.2cm, minimum height=0.85cm, anchor=center, inner sep=0pt] at ($(hidden_bot.west) + (-0.5*\horizontalsepsmall, \vertsepsmall)$) {};
	\node [draw, fill=black, thick, shape=rectangle, minimum width=0.2cm, minimum height=0.85cm, anchor=center, inner sep=0pt] at ($(hidden_bot.east)!0.5!(output_bot.west) + (0, \vertsepsmall)$) {};

\end{tikzpicture}}
  \caption{Example of a macro-pipeline architecture with two stages for a neural network with $N_l = 1$ hidden layers~\cite{Kurzo2018}. More macro-pipeline stages can be added to the pipeline to implement neural networks of arbitrary depth $N_l$.}\label{fig:pipeline-archi}
\end{figure}

As shown in the example of Fig.~\ref{fig:pipeline-archi}, in our architecture, the canceller NN layers are mapped to macro-pipeline stages. Each macro-pipeline stage requires several clock cycles to compute its outputs and it can start its computations as soon as valid outputs from the previous macro-pipeline stage become available. Due to the high throughput requirements of the SIC task, we instantiate one macro-pipeline stage for each layer in the NN that is used for cancellation.

\begin{figure}[t]
  \centering
	\scalebox{0.5}{\def\horizontalsep{4.9cm}
\def\vertsep{4.5cm}

\def\horizontalsepsmall{1.5cm}
\def\vertsepsmall{1.5cm}

\def\filterheight{1.54cm}
\def\filterwidth{2.5cm}

\begin{tikzpicture}[scale=0.65, thick]
  \tikzstyle{dspfilter}=[rectangle, draw=black, minimum width=\filterwidth, minimum height=\filterheight, inner sep=4pt, thick, align=center]
  \tikzstyle{bracket}=[decorate, decoration={brace, amplitude=5pt}]	
	%%%% NODES %%%%

	% Input Interface
	\node[dspfilter] (input_interface) at (0,0) {\textbf{Input} \\ \textbf{Interface}};

	% PE
	\coordinate (pe0_coord) at ($(input_interface.east)+(\horizontalsep, 0)$);
	\node[dspfilter] (pe2) at ($(pe0_coord)+(0.6, 0.6)$) {};
	\node[dspfilter, fill=white] (pe1) at ($(pe0_coord)+(0.3, 0.3)$) {};
	\node[dspfilter, fill=white] (pe0) at (pe0_coord) {\textbf{PE} \\ (internal \\ memory) };
	\draw[bracket] ($(pe0.north west) + (-0.2, -0.2)$) -- node[above left=0.2] {$N_{\text{PE}}$} ($(pe2.north west) + (-0.2, 0.2)$);	
	
	% Output Interface
	\node[dspfilter] (output_interface) at ($(pe0.east)+(\horizontalsep, 0)$) {\textbf{Output} \\ \textbf{Interface} \\ (tree adder, \\ act. function)};
	
	% Control Unit
	\node[dspfilter, minimum width=2.7*\filterwidth] (control_unit) at ($(input_interface.south)!0.5!(pe0.south)+(0, -\vertsep)$) {\textbf{Control Unit} \\ (counters, memory signal generation) \\ {} \\ {} \\ {} };	
	
	\node[ellipse, minimum width =0.75cm, minimum height=0.5cm, draw=red, thick] (state_left) at ($(control_unit)+(-1.0*\horizontalsepsmall, -0.75)$) {};	
	\node[ellipse, minimum width =0.75cm, minimum height=0.5cm, draw=red, thick] (state_right) at ($(control_unit)+(1.0*\horizontalsepsmall, -0.75)$) {};	
	
	\path[-triangle 45, thick, red]
		(state_left) edge[bend left=30] node [right] {} (state_right)
		(state_right) edge[bend left=30] node [left] {} (state_left);
	
	% Weight and bias memory
	\coordinate (w0_coord) at (control_unit.east -| output_interface.south);
	\node[dspfilter] (w2) at ($(w0_coord)+(0.6, 0.6)$) {};
	\node[dspfilter, fill=white] (w1) at ($(w0_coord)+(0.3, 0.3)$) {};
	\node[dspfilter, fill=white] (w0) at (w0_coord) {\textbf{Weight} \\ \textbf{and Bias} \\ \textbf{Memory}};

	%%%% CONNECTIONS %%%%
  
  	% Input Interface
	\draw[-triangle 45] ($(input_interface.west) + (-\horizontalsepsmall, 0)$) -- node [pos=-0.7] {{Data In}} ($(input_interface.west)$);  
	\draw[-triangle 45] (input_interface) -- (pe0);

	% Control Unit
	\draw[triangle 45-] ($(control_unit.west) + (-\horizontalsepsmall, 0)$) -- ($(control_unit.west)$);  
	\draw[-triangle 45] ($(control_unit.west) + (-\horizontalsepsmall, 0)$) -- ($(control_unit.west)$);  	
	\node[align=left] () at ($(control_unit.west) + (-1.75*\horizontalsepsmall, 0)$) {Pipeline \\ Control};
	
	\draw[triangle 45-] (input_interface) -- (control_unit.north -| input_interface);
	\node[align=right] () at ($(control_unit.north -| input_interface)!0.5!(input_interface.south) + (-1.5,0 )$) {Input \\ Selection};
		
	\draw[-triangle 45] (control_unit.north -| pe0) -- (pe0) ;	
	\node[align=left] () at ($(control_unit.north -| pe0)!0.5!(pe0.south) + (-1.5,0 )$) {PE Enable \\ Reset Sum};	
	
	\draw[-triangle 45] (control_unit) -- (w0);	
	\node[align=left] () at ($(control_unit.east)!0.5!(w0.west) + (0,-0.75 )$) {Memory \\ Signals};	
	
	%  Weight and bias
	\draw[-triangle 45] (w0.north west) -- (pe0.south east);		
	\node[align=left] () at ($(w0.north west)!0.5!(pe0.south east) + (0.75,0.25 )$) {Weights};		
	\draw[-triangle 45] (w0.north) -- node [pos=0.5, right] {Biases} (output_interface.south);			
	
	\draw[-triangle 45] ($(w0.east) + (\horizontalsepsmall, 0)$) -- ($(w0.east)$);  	
	\node[align=left] () at ($(w0.east) + (1.7*\horizontalsepsmall, 0)$) {External \\ Memory};		
	
	% PE
	\draw[-triangle 45] (pe0) -- (output_interface);		
	
	% Output Interface
	\draw[-triangle 45] (output_interface.east) -- node [pos=1.75] {Data Out} ($(output_interface.east) + (\horizontalsepsmall, 0)$);
	
	% Squares
	\node [draw, fill=black, thick, shape=rectangle, minimum width=0.2cm, minimum height=0.85cm, anchor=center, inner sep=0pt] at ($(pe0.east)!0.6!(output_interface.west)$) {};

\end{tikzpicture}}
  \caption{Block diagram of the macro-pipeline stage architecture~\cite{Kurzo2018}.}\label{fig:dense-layer-archi}
\end{figure}
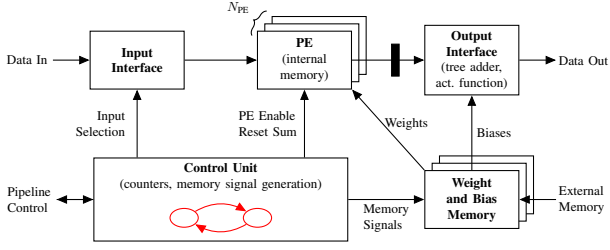

Let $\text{NE}_{l}$ denote the number of neurons in layer $l$. We note that $\text{NE}_0 = 2L$, $\text{NE}_{N_l+1} = 2$, and $\text{NE}_l = N_h$ for all hidden layers $l \in \{1,\hdots,\text{NE}_l\}$. The goal of a macro-pipeline stage is to compute $\mathbf{l}_{l}$ using expressions of the form~\eqref{eq:layer1}-\eqref{eq:layero}. Each element $j \in \{0,\hdots,{\text{NE}_l-1}\}$ of $\mathbf{l}_{l}$ can be computed as:
\begin{align}
  \mathbf{l}_{l}[j] = f_l \left( \mathbf{b}_l[j] + \sum_{i=0}^{\text{NE}_{l-1}-1} \mathbf{W}_l[i,j]\mathbf{l}_{l-1}[i] \right). \label{eq:dense-layer}
\end{align}
The architecture of each macro-pipeline stage is shown in more detail in Fig.~\ref{fig:dense-layer-archi}. More specifically, each macro-pipeline stage contains an input interface, an array of $N_{\text{PE}}$ processing elements (PEs), a weights-and-biases memory, a control unit, and an output interface. We note that for simplicity, all weights, biases, and partial sums have a common bit-width of $Q$ bits and saturation is used in case of an overflow. More sophisticated quantization schemes are possible, but they are beyond the scope of this work.

The $N_{\text{PE}}$ PEs, whose internal structure is shown in Fig.~\ref{fig:rpe-archi}, can be used to compute~\eqref{eq:dense-layer} over multiple clock cycles using one of two possible schedules. In the neuron-by-neuron (NBN) schedule, neurons are processed sequentially and each of the $N_{\text{PE}}$ PEs computes a part of the sum in~\eqref{eq:dense-layer} for a given neuron $j$. In the input-by-input (IBI) schedule, the inputs of layer $l$ (i.e., $\mathbf{l}_{l-1}$) are processed sequentially and the $N_{\text{PE}}$ PEs update the sum in~\eqref{eq:dense-layer} with the term $\mathbf{W}[i,j]\mathbf{l}_{l-1}[i]$ for $N_{\text{PE}}$ distinct neurons in parallel. 
As an NBN macro-pipeline stage generates neuron output values sequentially, the optimal accelerator structure consists of an NBN macro-pipeline stage always being followed by an IBI macro-pipeline stage, allowing the IBI stage to start performing computations once the output of the first neuron of the preceding NBN stage has been computed.
Once all inputs have been processed by the IBI stage, it immediately outputs multiple values to the NBN stage which follows it.
Having an NBN stage after another NBN stage means that the second NBN stage would have to wait for all outputs of the previous stage to be generated before any processing can take place, and having an IBI stage followed by another IBI stage would mean that the second IBI stage cannot start processing before the first IBI stage has processed all its inputs.
This structure of NBN and IBI stages, connected in an alternating fashion, masks a significant part of the latency and reduces the number of interconnects between two consecutive macro-pipeline stages. Since the exact architecture of each macro-pipeline stage depends on the processing schedule, we describe the details of the corresponding architectures separately in the next two sections.
\subsection{Neuron-by-Neuron Macro-Pipeline Architecture}
\subsubsection{Input Interface}
The input interface consists of $N_{\text{PE}}$ multiplexers, which route each of the $\text{NE}_{l-1}$ elements of $\mathbf{l}_{l-1}$ to the correct PE.
\subsubsection{Processing Elements}
In the NBN schedule, each PE is only associated with a single neuron, and therefore only a single partial sum needs to be stored in each PE. Thus, the PEs are simple multiply-and-accumulate (MAC) units and the memory shown in Fig.~\ref{fig:rpe-archi} is, in fact, a single $Q$-bit register.
\subsubsection{Control Unit}
The main tasks of the control unit are to distribute the computations to the PEs and to stall the computations when no valid inputs are available or when the following macro-pipeline stage is not ready to accept new inputs. The computations are dispatched to the PEs as follows. When $N_{\text{PE}} \leq \text{NE}_{l-1}$, all $N_{\text{PE}}$ PEs are used to process a single neuron at a time and $\text{NE}_l\left\lceil\frac{\text{NE}_{l-1}}{N_{\text{PE}}}\right\rceil$ clock cycles are required to process all neurons. When $N_{\text{PE}} > \text{NE}_{l-1}$, we constrain $N_{\text{PE}}$ so that $N_{\text{PE}} = k \cdot \text{NE}_{l-1},~k \in \mathbb{N},$ and hence $k$ neurons are processed in parallel and $\left\lceil\frac{\text{NE}_l\text{NE}_{l-1}}{N_{\text{PE}}}\right\rceil$ clock cycles are required to process all neurons.
\subsubsection{Weight and Bias Memories}
The weight and bias memories for layer $l$ are used to store $\mathbf{W}_l$ and $\mathbf{b}_l$ and they can be written externally to re-configure the NN canceller. The weights are organized in a memory that is $N_{\text{PE}}Q$ bits wide so that all PEs can be provided with data in parallel. A single word of the weight memory contains $N_{\text{PE}}$ weight values corresponding to $k$ different neurons. The bias memory, on the other hand, has a bit-width of $kQ$ bits.
\subsubsection{Output Interface}
The output interface adds the partial sums from the $N_{\text{PE}}$ PEs using an adder tree, it adds the corresponding biases, and it applies the non-linear activation function $f_l$ for each of the $k$ neurons that are being processed in parallel. A register is added between the PEs and the output interface to reduce the critical path of the architecture. Moreover, the output interface forwards the outputs of the $k$ neurons that are processed in parallel to the next macro-pipeline stage.
\subsubsection{Latency}
In the remainder of this work, we select $N_{\text{PE}}$ carefully so that both $\frac{\text{NE}_{l-1}}{N_{\text{PE}}}$ and $\frac{\text{NE}_l\text{NE}_{l-1}}{N_{\text{PE}}}$ are integers. With this setting, an NBN macro-pipeline stage requires
\begin{align}
 \mathcal{L}_l &  = \frac{\text{NE}_l\text{NE}_{l-1}}{N_{\text{PE}}} + 1 ,
\end{align}
clock cycles to produce all outputs of NN layer $l$. However, one full set of outputs for a NN layer is actually produced every $\frac{\text{NE}_l\text{NE}_{l-1}}{N_{\text{PE}}}$ cycles, so that the throughput of the NBN macro-pipeline stage in samples per clock cycle is
\begin{align}
 \mathcal{T}_l &  = \frac{N_{\text{PE}}}{\text{NE}_l\text{NE}_{l-1}}.
\end{align}
Moreover, the first $k$ outputs of an NBN macro-pipeline stage become available after
\begin{align}
 \mathcal{L}_{l,\text{first}} & = \left\lceil\frac{\text{NE}_{l-1}}{N_{\text{PE}}}\right\rceil + 1 ,
\end{align}
clock cycles. Therefore, a potential IBI macro-pipeline stage that follows can already start its computations after the $\mathcal{L}_\text{first}$ clock cycles and that only $k \leq \text{NE}_{l}$ outputs need to be forwarded to the next stage at a time.
\begin{figure}[t]
  \centering
	\scalebox{0.55}{\def\horizontalsep{3cm}
\def\vertsep{3.5cm}

\def\horizontalsepsmall{1.5cm}
\def\vertsepsmall{1.5cm}

\def\filterheight{4.5cm}
\def\filterwidth{1.75cm}

\def\muxinputsep{0.55cm}

\tikzset{
  multiplexer/.style={
    draw,
    trapezium, trapezium angle=67.5, draw,inner xsep=0pt,outer sep=0pt, minimum height=0.4cm, text width=0.65cm
  }  
}

\begin{tikzpicture}[-triangle 45, scale=0.65, thick]
  \tikzstyle{dspfilter}=[rectangle, draw=black, minimum width=\filterwidth, minimum height=\filterheight, inner sep=4pt, thick, align=center]

	%%%% NODES %%%%

	\node[circle, draw=black,inner sep=0pt, thick] (mult) at (0,0) {\Large $\bm{\times}$};

	\node[circle, draw=black,inner sep=0pt, thick] (add) at ($(mult.east)+ (\horizontalsep,0)$) {\Large $\bm{+}$};
	
	\node[multiplexer, rotate=90] (mux1) at ($(add.east)+ (\horizontalsepsmall,0)$) {};

	\node[multiplexer, rotate=-90] (mux2) at ($(add.east)+ (\horizontalsepsmall,-\vertsep)$) {};

	\node[dspfilter] (memory) at ($(add.east)+ (2.5*\horizontalsep,-\vertsepsmall)$) {\textbf{Memory}};

	%%%% CONNECTIONS %%%%

	% To Mult
	\draw[-triangle 45] ($(mult.west) + (-\horizontalsepsmall,0)$) -- node [pos=-0.75] {Data In} (mult);

	\draw[-triangle 45] ($(mult.west) + (-\horizontalsepsmall,-\vertsepsmall)$) -| node [pos=-0.35] {Weight In} (mult.south);

	% To Adder
	\draw[-triangle 45] (mult) -- (add);
	\draw[-triangle 45] (mux1.north) -- (add);

	% To Mux 1
	\draw[-triangle 45] ($(mux1.south) + (\horizontalsepsmall,\muxinputsep)$) -- node [pos=-0.25] {0} ($(mux1.south)+(0, \muxinputsep)$);

	\coordinate (mux1_mem) at (mux1.south -| memory.west);
	\coordinate (mux1_mem_p1) at ($(mux1_mem) + (0,-\muxinputsep)$);
	\coordinate (mux1_mem_p2) at ($(mux1.south)+(0, -\muxinputsep)$);
	\draw[-triangle 45] (mux1_mem_p1) -- node [above, pos=0.5] {Partial Sum} (mux1_mem_p2);
	\node[thick, circle, fill=black, minimum width=0.175cm, inner sep=0] at ($(mux1_mem_p1)!0.5!(mux1_mem_p2)$) {};			

	\node[] at ($(mux1.south) + (-0.315cm,\muxinputsep)$) {$1$};	
	\node[] at ($(mux1.south) + (-0.315cm,-\muxinputsep)$) {$0$};

	% To Mux 2
	\draw[-triangle 45] (add) |- ($(mux2.south) + (0,-\muxinputsep)$);
	
\draw[-triangle 45] ($(mux1_mem_p1)!0.5!(mux1_mem_p2)$) |- ($(mux2.south) + (-1.0cm,3*\muxinputsep)$) |- ($(mux2.south) + (0,\muxinputsep)$);

	\node[] at ($(mux2.south) + (0.315cm,-\muxinputsep)$) {$1$};
	\node[] at ($(mux2.south) + (0.315cm,\muxinputsep)$) {$0$};

	% To memory
	
	\draw[-triangle 45] ($(memory.east) + (1.2*\horizontalsepsmall,0)$) -- node[align=center] [pos=-0.65] {Memory \\ Interface} (memory.east);	
	
	\coordinate (mux2_mem) at ($(mux2)!0.5!(mux2 -| memory.west)$);	
	\draw[-triangle 45] (mux2) -- (mux2 -| memory.west);	
	\draw[-triangle 45] (mux2_mem) |- node [pos=1.09] {Data Out} ($(mux2_mem) + (2.25*\horizontalsep, -0.75*\vertsep)$);		

	\node[thick, circle, fill=black, minimum width=0.175cm, inner sep=0] at (mux2_mem) {};			

	% Box around it all
	\draw[thin] (-0.9*\horizontalsepsmall,-2*\vertsep) rectangle (4.6*\horizontalsep,0.75*\vertsep);

\end{tikzpicture}}
  \caption{Detailed view of the PE architecture that is used by both the NBN and the IBI macro-pipeline stages~\cite{Kurzo2018}.}\label{fig:rpe-archi}
\end{figure}
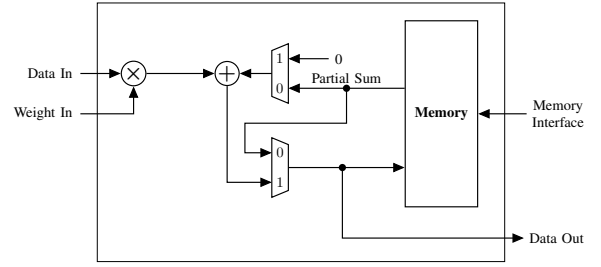
\subsection{Input-by-Input Macro-Pipeline Architecture}
\subsubsection{Input \& Output Interfaces}
The input and output interfaces of the IBI macro-pipeline stage are similar to that of the NBN macro-pipeline stage. The main difference is that the IBI output interface forwards the outputs of all $\text{NE}_l$ neurons that are processed in parallel to the next macro-pipeline stage.
\subsubsection{Processing Elements}
In the IBI schedule, each PE can be associated with multiple neurons. Therefore, several partial sums may need to be stored in each PE. Thus, the PEs are MAC units and the memory shown in Fig.~\ref{fig:rpe-archi} has $\left\lceil\frac{\text{NE}_l}{N_{\text{PE}}}\right\rceil Q$ bits.
\subsubsection{Control Unit}
In the IBI schedule, when $N_{\text{PE}} \leq \text{NE}_l$, all $N_{\text{PE}}$ PEs are used to update the $\text{NE}_l$ neurons of layer $l$ sequentially with a new input value $\mathbf{l}[i]$ and $\text{NE}_{l-1}\left\lceil\frac{\text{NE}_l}{N_{\text{PE}}}\right\rceil$ clock cycles are required to process all neurons. When $N_{\text{PE}} > \text{NE}_l$, we constrain $N_{\text{PE}}$ so that $N_{\text{PE}} = k\text{NE}_l,~k \in \mathbb{N},$ and $k$ inputs are processed in parallel. Hence, $\left\lceil\frac{\text{NE}_l\text{NE}_{l-1}}{N_{\text{PE}}}\right\rceil$ clock cycles are required to process all neurons.
\subsubsection{Weight and Bias Memories}
The weight and bias memories are similar to those of the NBN macro-pipeline stage. A single word of the weight memory contains $N_{\text{PE}}$ weights corresponding to $k$ different neurons. The bias memory has a bit-width of $\text{NE}_l Q$ bits in the IBI macro-pipeline stage. All memories support external writes to re-configure the canceller.
\subsubsection{Latency}
Similarly to the NBN schedule, we choose $N_{\text{PE}}$ carefully so that both $\frac{\text{NE}_l}{N_{\text{PE}}}$ and $\frac{\text{NE}_l\text{NE}_{l-1}}{N_\text{PE}}$ are always integers. Then, the latency and the throughput are
\begin{align}
 \mathcal{L}_l &  = \frac{\text{NE}_l\text{NE}_{l-1}}{N_{\text{PE}}} + 1,
\end{align}
clock cycles and 
\begin{align}
 \mathcal{T} & = \frac{N_{\text{PE}}}{\text{NE}_l\text{NE}_{l-1}},
\end{align}
samples per clock cycle, respectively. Moreover, since all $\text{NE}_l$ outputs of an IBI macro-pipeline stage become available simultaneously, the number of clock cycles until the first output is identical to $\mathcal{L}_l$ and also given by
\begin{align}
 \mathcal{L}_{l,\text{first}} & =  \frac{\text{NE}_l\text{NE}_{l-1}}{N_{\text{PE}}} + 1 ,
\end{align}
clock cycles.
\subsection{Overall Neural Network Canceller Architecture}
The overall NN architecture consists of $N_l$ macro-pipeline stages with pipeline registers added between them. The first hidden layer uses an NBN macro-pipeline stage and the second hidden layer (or the output layer when $N_l=1$) uses an IBI macro-pipeline stage. Further layers use NBN and IBI macro-pipeline stages in an alternating fashion as previously discussed. The $\text{NE}_0 = 2L$ inputs $\mathbf{l}_0$ of the first NBN macro-pipeline stage that implements the computations of the first hidden layer are assumed to all be available in parallel. The number of PEs instantiated for layer $l$ is denoted by $N_{\text{PE},l}$. The computations for the linear canceller are done in parallel with the NN by instantiating a standard complex FIR filter with $N_{\text{PE},\text{linear}}$ complex-valued PEs. The latency of the linear canceller in clock cycles
\begin{align}
	\mathcal{L}_{\text{linear}}	& = \left\lceil \frac{L}{N_{\text{PE},\text{linear}}} \right\rceil.
\end{align}
Since the linear canceller is not pipelined, it holds that $\mathcal{T}_{\text{linear}}~=~ \frac{1}{\mathcal{L}_{\text{linear}}}$. The throughput of the overall NN canceller architecture is:
\begin{align}
  \mathcal{T} & = \min \left\{\mathcal{T}_{\text{linear}}, \; \min_{l\in\{1,\hdots,N_l+1\}} \mathcal{T}_l\right\}.
\end{align}
Since it is typically not very costly in terms of resources to ensure that $\mathcal{T}_{\text{linear}} \geq \mathcal{T}_l,~l\in\{1,\hdots,N_{l+1}\},$ in practice $\mathcal{T}$ is usually limited by $\min_{l} \mathcal{T}_l$. As opposed to the throughput, the latency of the overall NN canceller is more complicated to derive in general. However, in the special case where the number of PEs for each layer $l$ is chosen such that no stalling happens and $N_l+1$ is even, the latency can be calculated as:
\begin{align}
 \mathcal{L} & =  \max \left\{\mathcal{L}_{\text{linear}}, \; \sum _{l=1}^{(N_l+1)/2}\left(\mathcal{L}_{2l-1,\text{first}}+\mathcal{L}_{2l}\right)\right\}, \label{eq:latnn}
\end{align}
where the odd-indexed terms in the summation correspond to NBN macro-pipeline stages and the even terms correspond to IBI macro-pipeline stages.
Finally, we note that the denormalization step shown in Fig.~\ref{fig:si-canceller-nn} is constrained to scaling with powers of two, which can be implemented efficiently with simple shifting operations, both during training and during inference.
\begin{figure}[t]
  \centering
	\scalebox{0.5}{\def\horizontalsep{4.9cm}
\def\vertsep{4.5cm}

\def\horizontalsepsmall{1.5cm}
\def\vertsepsmall{1.5cm}

\def\filterheight{1.54cm}
\def\filterwidth{2.5cm}

\begin{tikzpicture}[scale=0.65, thick]
  \tikzstyle{dspfilter}=[rectangle, draw=black, minimum width=\filterwidth, minimum height=\filterheight, inner sep=4pt, thick, align=center]
  \tikzstyle{bracket}=[decorate, decoration={brace, amplitude=5pt}]	
	%%%% NODES %%%%

	% Input Interface
	\node[dspfilter] (input_interface) at (0,0) {\textbf{Basis} \\ \textbf{Functions} \\ \textbf{Calculator/} \\ \textbf{Memory}};

	% PE
	\coordinate (pe0_coord) at ($(input_interface.east)+(\horizontalsep, 0)$);
	\node[dspfilter] (pe2) at ($(pe0_coord)+(0.6, 0.6)$) {};
	\node[dspfilter, fill=white] (pe1) at ($(pe0_coord)+(0.3, 0.3)$) {};
	\node[dspfilter, fill=white] (pe0) at (pe0_coord) {\textbf{Complex PE} \\ (single register) };
	\draw[bracket] ($(pe0.north west) + (-0.2, -0.2)$) -- node[above left=0.2] {$N_{\text{CPE}}$} ($(pe2.north west) + (-0.2, 0.2)$);	
	
	% Output Interface
	\node[dspfilter] (output_interface) at ($(pe0.east)+(\horizontalsep, 0)$) {\textbf{Output} \\ \textbf{Block} \\ (tree adder)};
	
	% Control Unit
	\node[dspfilter, minimum width=2.7*\filterwidth] (control_unit) at ($(input_interface.south)!0.5!(pe0.south)+(0, -\vertsep)$) {\textbf{Control Unit} \\ (counters, memory signal generation) \\ {} \\ {} \\ {} };	
	
	\node[ellipse, minimum width =0.75cm, minimum height=0.5cm, draw=red, thick] (state_left) at ($(control_unit)+(-1.0*\horizontalsepsmall, -0.75)$) {};	
	\node[ellipse, minimum width =0.75cm, minimum height=0.5cm, draw=red, thick] (state_right) at ($(control_unit)+(1.0*\horizontalsepsmall, -0.75)$) {};	
	
	\path[-triangle 45, thick, red]
		(state_left) edge[bend left=30] node [right] {} (state_right)
		(state_right) edge[bend left=30] node [left] {} (state_left);
	
	% Weight and bias memory
	\coordinate (w0_coord) at (control_unit.east -| output_interface.south);
	\node[dspfilter] (w2) at ($(w0_coord)+(0.6, 0.6)$) {};
	\node[dspfilter, fill=white] (w1) at ($(w0_coord)+(0.3, 0.3)$) {};
	\node[dspfilter, fill=white] (w0) at (w0_coord) {\textbf{Complex} \\ \textbf{Weight} \\ \textbf{Memory}};

	%%%% CONNECTIONS %%%%
  
  	% Input Interface
	\draw[-triangle 45] ($(input_interface.west) + (-\horizontalsepsmall, 0)$) -- node [pos=-0.7] {{Data In}} ($(input_interface.west)$);  
	\draw[-triangle 45] (input_interface) -- (pe0);

	% Control Unit
	\draw[triangle 45-] ($(control_unit.west) + (-\horizontalsepsmall, 0)$) -- ($(control_unit.west)$);  
	\draw[-triangle 45] ($(control_unit.west) + (-\horizontalsepsmall, 0)$) -- ($(control_unit.west)$);  	
	\node[align=left] () at ($(control_unit.west) + (-1.75*\horizontalsepsmall, 0)$) {Pipeline \\ Control};
	
	\draw[triangle 45-] (input_interface) -- (control_unit.north -| input_interface);
	\node[align=right] () at ($(control_unit.north -| input_interface)!0.5!(input_interface.south) + (-2.0,0 )$) {Basis Function \\ Selection};
		
	\draw[-triangle 45] (control_unit.north -| pe0) -- (pe0) ;	
	\node[align=left] () at ($(control_unit.north -| pe0)!0.5!(pe0.south) + (-1.5,0 )$) {PE Enable \\ Reset Sum};	
	
	\draw[-triangle 45] (control_unit) -- (w0);	
	\node[align=left] () at ($(control_unit.east)!0.5!(w0.west) + (0,-0.75 )$) {Memory \\ Signals};	
	
	%  Weight and bias
	\draw[-triangle 45] (w0.north west) -- (pe0.south east);		
	\node[align=left] () at ($(w0.north west)!0.5!(pe0.south east) + (0.75,0.25 )$) {Weights};		
	\draw[-triangle 45] (w0.north) -- node [pos=0.5, right] {Biases} (output_interface.south);			
	
	\draw[-triangle 45] ($(w0.east) + (\horizontalsepsmall, 0)$) -- ($(w0.east)$);  	
	\node[align=left] () at ($(w0.east) + (1.7*\horizontalsepsmall, 0)$) {External \\ Memory};		
	
	% PE
	\draw[-triangle 45] (pe0) -- (output_interface);		
	
	% Output Interface
	\draw[-triangle 45] (output_interface.east) -- node [pos=1.75] {Data Out} ($(output_interface.east) + (\horizontalsepsmall, 0)$);
	
	% Squares
	\node [draw, fill=black, thick, shape=rectangle, minimum width=0.2cm, minimum height=0.85cm, anchor=center, inner sep=0pt] at ($(pe0.east)!0.6!(output_interface.west)$) {};

\end{tikzpicture}}
  \caption{Block diagram of the polynomial canceller architecture.}\label{fig:poly-archi}
\end{figure}
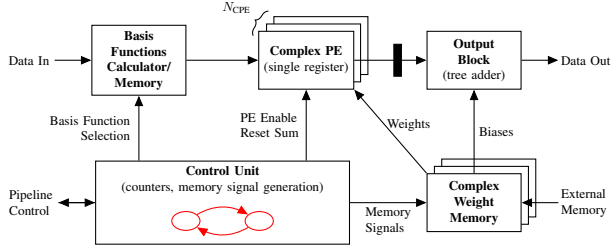
\section{Polynomial Canceller Hardware Architecture}\label{sec:polyhardware}
Since, to the best of our knowledge, there are no published implementations of polynomial SI cancellers in the literature, we provide our own optimized reference implementation. Our polynomial SI canceller architecture, which is shown in Fig.~\ref{fig:poly-archi}, is largely based on the NN architecture since the main computational tasks of the two cancellers are very similar (i.e., computation of weighted sums). The main differences are that the input interface also computes the basis functions,  that $N_{\text{CPE}}$ complex PEs (CPEs) are used to perform computations on complex values, and that there is only a single macro-pipeline stage. In the remainder of this section, we explain how the basis functions can be computed efficiently and we describe the polynomial SI canceller in more detail.
\subsection{Basis Function Computation}
The computation of the $N_{\text{BF}}$ basis functions in \eqref{eq:basisfunction} for each cancellation sample seems like a cumbersome task. Fortunately, we can show that the basis functions have a number of properties that enable their efficient computation. First, significant basis function re-use is possible. In particular, after $\hat{y}_{\text{SI}}[n-1]$ has been computed based on $\text{BF}_{p,q}(x[n{-}1{-}l]),~l\in \{0,\hdots,{L{-}1}\},~p\in \{1,3,\hdots,P\},~q \in \{0,\hdots,p\}$, the basis functions for $l\in \{0,\hdots,{L{-}2}\}$ can be stored and re-used for the computation of $\hat{y}_{\text{SI}}[n]$. As such, the only new basis functions that need to be computed for $\hat{y}_{\text{SI}}[n]$ are $\text{BF}_{p,q}(x[n]),~p\in \{1,3,\hdots,P\},~q \in \{0,\hdots,p\}$. This requires $\frac{L-1}{4}\left(P+1\right)\left(P+3\right)$ memory elements, but reduces the number of basis functions that need to be computed by a factor of $L$ from $\frac{L}{4}(P+1)(P+3)$ to $\frac{1}{4}(P+1)(P+3)$. Moreover, the following proposition shows two additional properties of the basis functions.
\begin{algorithm}[t]
  \caption{Dynamic programming computation of basis functions $\text{BF}_{p,q}(x[n])$}
  \label{alg:bf}
  \begin{algorithmic}[1]\small
    \State \textbf{Input:} $x[n]$
    \State \textbf{Outputs:} $\text{BF}_{p,q}(x[n])$ for $p \in \{1,3,\hdots,P\},~q\in\{0,\hdots,p\}$
	\State $\text{BF}_{1,0}(x[n]) \gets (x[n])^*$
	\State $\text{BF}_{1,1}(x[n]) \gets x[n]$
	\For{$p \in \{3,5,\hdots,P\}$}
	\For{$q \in \left\{\frac{p+1}{2},\hdots,p\right\}$}
		\State $\text{BF}_{p,q}(x[n]) \gets \left(x[n]\right)^2\text{BF}_{p-2,q-2}(x[n])$ \label{alg:bf:mult}
		\State $\text{BF}_{p,p-q}(x[n]) \gets \text{BF}_{p,q}(x[n])^*$ \label{alg:bf:conj}
	\EndFor
	\EndFor
  \end{algorithmic}
\end{algorithm}
\begin{proposition}
For the basis functions in \eqref{eq:basisfunction}, it holds that:
\begin{enumerate}
	\item $\text{BF}_{p,q}(x) = \left(\text{BF}_{p,p-q}(x)\right)^*$
	\item $\text{BF}_{p,q}(x) = x^2\text{BF}_{p-2,q-2}(x)$
\end{enumerate}
\end{proposition}
\begin{IEEEproof}
Both properties follow from the definition of the basis function in~\eqref{eq:basisfunction}. Specifically, for 1) we have:
\begin{align}
	\text{BF}_{p,q}(x)		& = x^{q}(x^*)^{p-q} \nonumber \\
							& = \left(x^{p-q}(x^*)^{p-(p-q)}\right)^* \nonumber \\
							& = \left(\text{BF}_{p,p-q}(x)\right)^*,
\end{align}
and for 2) we have:
\begin{align}
	\text{BF}_{p,q}(x) 		& = x^q(x^*)^{p-q}  \nonumber\\
							& = x^2x^{q-2}(x^*)^{p-2-(q-2)} \nonumber \\
							& = x^2\text{BF}_{p-2,q-2}(x).
\end{align}
\end{IEEEproof}
Property 1) enables a computation reduction by a factor of two since for every $p \in \{1,3,\hdots,P\}$, it is sufficient to compute $\text{BF}_{p,q}(x)$ only for $q \in \left\{\frac{p+1}{2},\hdots,p\right\}$ and the remaining basis functions for $q \in \left\{0,\hdots,\frac{p-1}{2}\right\}$ can be obtained by simple conjugation. Moreover, property 2) reveals an efficient dynamic programming (DP) method to compute the basis functions for $x[n]$, which is shown in Algorithm~\ref{alg:bf}. Algorithm~\ref{alg:bf} requires one multiplication to pre-compute $(x[n])^2$ and $\frac{1}{8}(P+1)(P+3)-2$ multiplications for all executions of line~\ref{alg:bf:mult}. The conjugation in line~\ref{alg:bf:conj} does not require any multiplications as it is a simple sign change of the imaginary part of $\text{BF}_{p,p-q}(x)$. As such, the total number of multiplications to compute the basis functions for a baseband sample $x[n]$ is:
\begin{align}
	N_{\text{MUL},\text{BF}}	& = \frac{1}{8}(P+1)(P+3)-1.
\end{align}
One downside of the DP approach is that only the inner loop in Algorithm~\ref{alg:bf} can be parallelized. However, in most typical applications we have $P \leq 9$, so that the outer loop in Algorithm~\ref{alg:bf} is executed very few times. We note that, due to the efficiency of Algorithm~\ref{alg:bf}, $N_{\text{MUL},\text{BF}}$ is significantly smaller than $N_{\text{MUL},\text{poly}}$, which justifies ignoring the multiplications of the basis function computations in \eqref{eq:multpoly} for simplicity.
\subsection{Polynomial Canceller Architecture}
We use a high-level structure that is similar to the NN-based cancellers in Fig.~\ref{fig:si-canceller-nn} in the sense that linear cancellation is done in parallel to non-linear cancellation and the polynomial SI canceller focuses only on the non-linear part of the SI signal. Since most of the SI signal is linear, removing the linear term separately significantly reduces the dynamic range of the values within the polynomial SI canceller, which in turn allows us to reduce the common quantization bit-width $Q$ for the real and the imaginary parts of the involved quantities.
\subsubsection{Input \& Output Interfaces}
The input interface consists of $N_{\text{CPE}}$ multiplexers, which route each of the $N_{\text{BF}}$ BFs to the correct CPE to compute parts of the sum in~\eqref{eq:finalbf}. As mentioned previously, the input interface also computes the BFs using $N_{\text{CPE},\text{BF}}$ CPEs. Since only the inner loop in Algorithm~\ref{alg:bf} can be parallelized, it is reasonable to constrain $N_{\text{CPE},\text{BF}}$ so that $N_{\text{CPE},\text{BF}} \leq \frac{P+1}{2}$. The number of clock cycles to compute all new BFs based on $x[n]$ with $N_{\text{CPE},\text{BF}}$ PEs is:
\begin{align}
	\mathcal{L}_{\text{BF,new}}	& = 1+\sum _{\substack{p=3,\\p \text{ odd}}}^{P} \left\lceil\frac{p+1}{2N_{\text{CPE},\text{BF}}}\right\rceil, \label{eq:latencybf}
\end{align}
where one clock cycle is used to pre-compute $x^2$ and the result (as well as $x^*$) are stored in two $2Q$-bit registers. The $\frac{L-1}{4}\left(P+1\right)\left(P+3\right)$ BFs that are re-used are stored in a circular buffer.
The output interface consists of an adder tree for the partial sums stored in the $N_{\text{CPE}}$ CPEs to produce the final result.
\subsubsection{Complex Processing Elements}
The $N_{\text{CPE}}$ CPEs are complex MAC units with a $Q$-bit register to store partial sums. The complex MAC units are implemented using three real-valued multipliers and five real-valued adders. 
\subsubsection{Control Unit}
Similarly to the NN-based canceller, the main tasks of the control unit are to distribute the computations to the CPEs and to stall the computations when no valid inputs are available. The control unit schedules the operation so that the CPEs first compute the terms of~\eqref{eq:finalbf} that are based on BFs that are already available in the circular buffer. In the meantime, the input interface computes the $\frac{1}{4}(P+1)(P+3)$ BFs that depend on the new sample $x[n]$.
\subsubsection{Parameter Memory}
The parameter memory is used to store the complex-valued $\hat{h}_{p,q}$ parameters of the polynomial canceller. The memory contains $\left\lceil\frac{N_{\text{BF}}}{N_{\text{CPE}}}\right\rceil$ words that are $2QN_{\text{CPE}}$ bits wide so that all $N_{\text{CPE}}$ CPEs can be provided with the parameters in parallel. The parameter memory can be written to externally to re-configure the polynomial canceller.
\subsubsection{Latency}
The terms of~\eqref{eq:finalbf} that are based on BFs and are available in the circular buffer can be computed in parallel to the computation of the new BFs that are based on $x[n]$, masking a part of the latency of the computation of~\eqref{eq:finalbf} or the new BFs (whichever is greater). The latency of computing the terms of~\eqref{eq:finalbf} for the BFs available in the circular buffer is:
%\new{The terms of~\eqref{eq:finalbf} that are based on BFs and are available in the circular buffer can be computed in parallel to the computation of the new BFs that are based on $x[n]$. As such, a part of the latency of the computation of~\eqref{eq:finalbf} or the new BFs (whichever is greater) can be masked. The latency of computing the terms of~\eqref{eq:finalbf} for the BFs that are available in the circular buffer is:
\begin{align}
	\mathcal{L}_{\text{BF,old}}	& = \left\lceil\frac{L-1}{L} \frac{N_{\text{BF}}}{N_{\text{CPE}}}\right\rceil.
\end{align}
Then, it can be shown that the overall latency of the polynomial canceller is given by:
\begin{align}
	\mathcal{L}_{\text{poly}}	& = 
	\left\{ 
		\begin{array}{ll} 
			\left\lceil\frac{N_{\text{BF}}}{N_{\text{CPE}}}\right\rceil+1, & \mathcal{L}_{\text{BF,old}} \geq \mathcal{L}_{\text{BF,new}}, \\
			\mathcal{L}_{\text{BF,new}} + \left\lceil\frac{1}{L} \frac{N_{\text{BF}}}{N_{\text{CPE}}}\right\rceil + 1, & \mathcal{L}_{\text{BF,old}} < \mathcal{L}_{\text{BF,new}},		
		\end{array} 
	\right.
\end{align}
where one clock cycle is required by the adder tree in the output interface to produce the final output. Since a pipeline register is inserted before the adder tree of the output interface, the throughput of the polynomial SI canceller, measured in samples per clock cycle, is given by:
\begin{align}
	\mathcal{T}_{\text{poly}} & = \frac{1}{\mathcal{L}_{\text{poly}}-1}. \label{eq:latnncanc}
\end{align}
\begin{figure}[t]
  \centering
  \includegraphics[trim={0 5cm 0 5cm},clip, width=0.925\columnwidth]{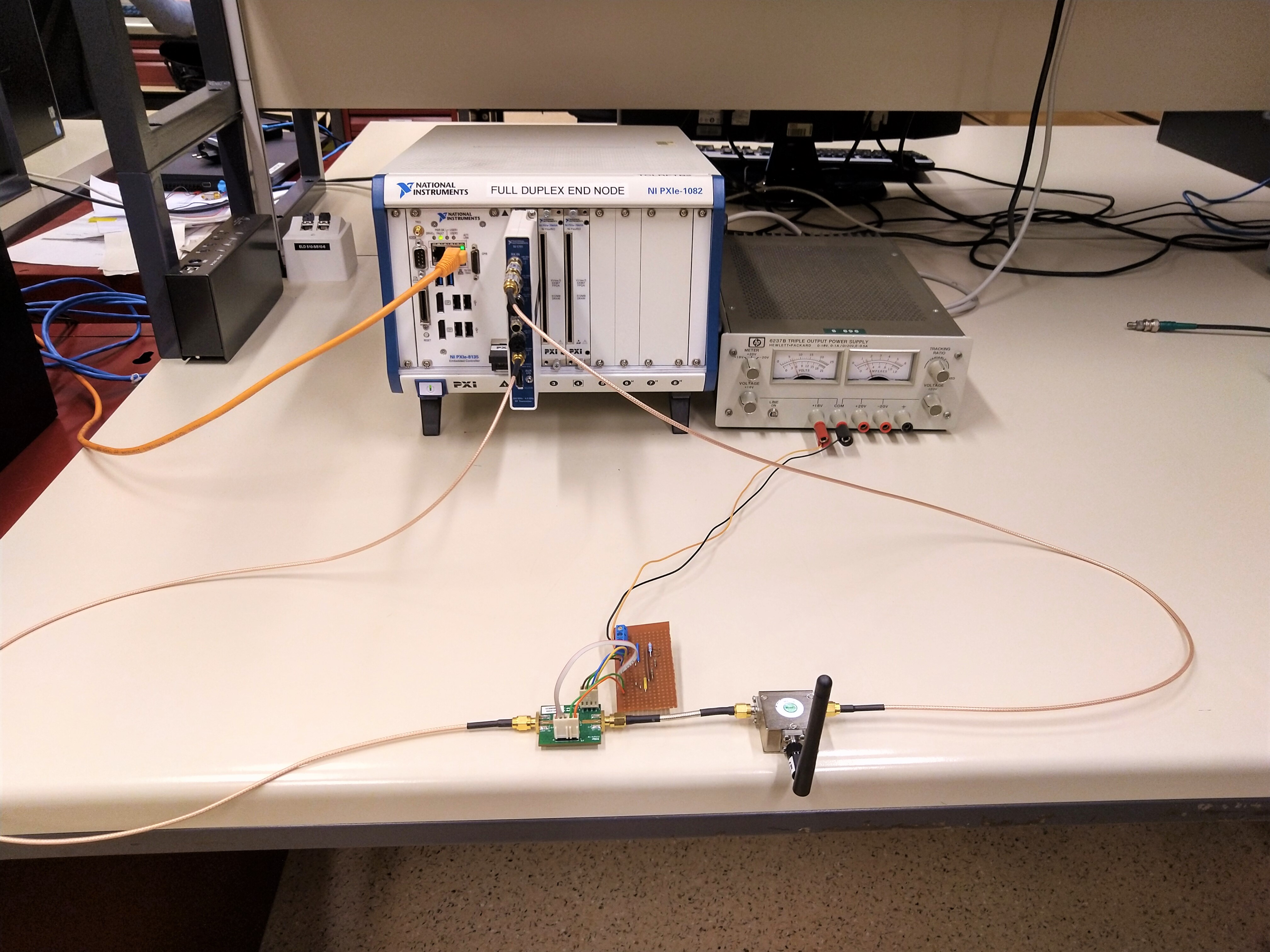}
  \caption{Our full-duplex testbed based on the National Instruments PXI platform with a NI 5791 RF card, a Skyworks SE2576L PA, a MECA CS-2.500 circulator, and $50$~dB of attenuation at the receiver to emulate active RF cancellation.}\label{fig:testbed}
\end{figure}
\section{Self-Interference Cancellation Results}\label{sec:sicresults}
In this section, we compare the polynomial SI canceller with the NN-based SI cancellers in terms of their SIC performance and their complexity. To this end, we first describe our full-duplex testbed and the employed dataset in detail. Then, we provide a high-level performance and complexity comparison of the polynomial SI canceller with the NN SI canceller.

\subsection{Full-Duplex Testbed}
A picture of our full-duplex testbed is shown in Fig.~\ref{fig:testbed}.
Our full-duplex testbed that is used to generate the dataset in this section is based on the National Instruments PXI platform~\cite{NIPXI} with a National Instruments NI-5791 RF card~\cite{NI5791}. The NI-5791 RF card is used for both transmission and reception. The transmitter and the receiver are configured to use the same local RF oscillator for up-conversion and down-conversion, respectively. The built-in IQ imbalance compensation is disabled to be consistent with commercial off-the-shelf transceivers. The NI-5791 RF card uses a $16$-bit Texas Instruments DAC3482 DAC~\cite{TIDAC3482} and a $14$-bit Texas Instruments ADS4246 ADC~\cite{TIADS4246}. As the NI-5791 RF card has a relatively low maximum output power\footnote{All output powers in this section refer to \emph{peak} output powers.} of $10$~dBm, we use an external Skyworks SE2576L PA~\cite{SkyworksSE2576L}. We use a MECA {CS-2.500} circulator~\cite{MECA} to provide approximately $15$~dB of isolation between the transmitter and the receiver.\footnote{The circulator datasheet~\cite{MECA} claims that the typical isolation is $20$~dB. This is true when the terminal where the antenna would be connected is terminated using a $50~\Omega$ terminator. However, when we connect a standard $2.4$~GHz whip antenna the measured isolation drops to $15$~dB due to reflections caused by imperfect impedance matching.} Since we do not have access to an active RF canceller, we emulate the active RF cancellation using a $50$ dB attenuator before the receiver input of the NI-5791 RF card. We note that this is a  feasible amount of active RF cancellation, which even the very first RF cancellers were able to achieve~\cite{Bharadia2013}. The model number, gain, and noise figure of the receiver LNA are not stated explicitly in~\cite{NI5791}, but the overall receiver noise figure is guaranteed to be less than $8$~dB at a frequency of $2$~GHz.
\subsection{Full-Duplex Dataset}
The transmitted signal is a $20$~MHz QPSK-modulated OFDM signal with $2048$ carriers and a peak-to-average power ratio (PAPR) of $13$~dB. The output power of the NI-5791 RF card is experimentally set so that the Skyworks SE2576L PA operates at its $1$~dB compression point, namely at an  output power of approximately $32$~dBm~\cite{SkyworksSE2576L}. The RF carrier frequency is set to $2.45$~GHz and the sampling rate of the receiver is set to $80$~MHz so that we oversample the OFDM signal by a factor of $4$. The dataset contains $20\,480$ time-domain SI baseband samples, out of which 90\% is used for training and 10\% for the evaluation of the SIC performance. For NN training, we use a mini-batch size of $B=32$ and the Adam optimizer~\cite{Kingma2015} with a learning rate of $\lambda = 0.004$.

One important issue is to ensure that our dataset is not obtained in a regime where we are limited by transmitter or receiver quantization noise. The $16$-bit DAC has a dynamic range of approximately $96$~dBm, which puts the transmitter quantization noise at approximately $-64$~dBm for a $32$~dBm transmit power. The $65$~dB of total isolation between the transmitter and the receiver attenuates all components of the signal equally. Thus, the transmitter quantization noise power at the receiver is approximately $-129$~dBm, which is well below the $-95$~dBm thermal noise power ($25\degree$~C, $80$~MHz bandwidth). The power at the LNA input is $-45.6$~dBm and the reference level of the receiver is set at its lowest supported value of $-27$~dBm. The $14$-bit DAC has a dynamic range of approximately $84$~dBm. As such, the receiver quantization noise floor is located at approximately $-111$~dBm, which is also well below the thermal noise power.
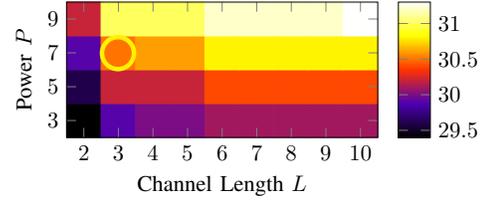
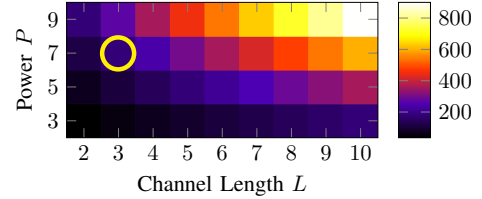
\begin{figure}[t]
	\centering
	\begin{subfigure}{0.9\linewidth}
		\centering
		\scalebox{0.85}{\begin{tikzpicture}

\begin{axis}[
	width=6.45cm,
	height = 3.7cm,
    ylabel=Power $P$,
    xlabel=Channel Length $L$,
    xtick distance=1,
    ytick={3,5,7,9},
    colormap/thermal,
    colorbar,
    disabledatascaling, 
    enlargelimits=false,
    axis on top,
    ]
   \addplot[matrix plot*,point meta=explicit] file [meta=index 2] {fig/data/heatmap_poly_perf.dat};

\end{axis}

\draw[line width=2, yellow] (0.805, 1.32) circle (0.25); 
\end{tikzpicture}%
}
		\caption{SIC performance in dB.}
		\label{fig:heatmap_poly_perf}
	\end{subfigure}
	\vskip\baselineskip
	\begin{subfigure}{0.9\linewidth}
		\centering
		\scalebox{0.85}{\begin{tikzpicture}

\begin{axis}[
	width=6.45cm,
	height = 3.7cm,
    ylabel=Power $P$,
    xlabel=Channel Length $L$,
    xtick distance=1,
    ytick={3,5,7,9},
    colormap/thermal,
%    colormap name=mymap3,
    colorbar,
    disabledatascaling, 
    enlargelimits=false,
    axis on top,
    ]
   \addplot[matrix plot*,point meta=explicit] file [meta=index 2] {fig/data/heatmap_poly_complexity.dat};
\end{axis}
\draw[line width=2, yellow] (0.805, 1.32) circle (0.25); 
\end{tikzpicture}%
}
		\caption{Number of multiplications.}
		\label{fig:heatmap_poly_complexity}
	\end{subfigure}
\caption{SIC performance and number of multiplications for the polynomial canceller as a function of the channel length $L$ and the maximum power $P$. The selected canceller with $L=3$ and $P=7$ is marked with yellow circles.}
\label{fig:heatmap_poly}
\end{figure}
\subsection{Comparison Setup}\label{sec:comparison}
The complexity expressions for the polynomial SI canceller in~\eqref{eq:addpoly}-\eqref{eq:multpoly} and the NN SI cancellers in~\eqref{eq:addnn}-\eqref{eq:multnn} cannot be compared directly because they contain different sets of parameters and they have different SIC performance. Thus, we choose to compare two pairs of points in the design space of the polynomial and the NN cancellers:
\begin{enumerate}
	\item A pair of points where the polynomial and the NN cancellers achieve their maximum respective SIC performance (\emph{peak-performance}).
	\item A pair of points where the polynomial and the NN cancellers have approximately the same SIC performance (\emph{equi-performance}).
\end{enumerate}
The comparison points are selected as follows. First, we evaluate the SIC performance of the polynomial canceller for various combinations of $L$ and $P$. Then, we find the maximum SIC performance $C_{\max,\text{poly}}$ and we select the combination of $L$ and $P$ that results in the smallest number of multiplications according to~\eqref{eq:multpoly} and has SIC performance at most $1$~dB lower than the maximum SIC performance. The back-off of $1$~dB is allowed because as $L$ and $P$ are increased there are severely diminishing returns in terms of the SIC performance and we want to ensure that a reasonable complexity-performance trade-off point is selected to be fair to the polynomial canceller. This gives the peak-performance point for the polynomial canceller. For the peak-performance point of the NN canceller, we follow the same procedure for various values of $L$, $N_h$, and $N_l$ and using~\eqref{eq:multnn} for the complexity evaluation. For the equi-performance point of the NN canceller, we select the NN with the smallest number of multiplications that achieves SIC performance greater than or equal to $C_{\max,\text{poly}}$.

In Fig.~\ref{fig:heatmap_poly}, we show two heatmaps for the SIC performance and the number of multiplications for the polynomial canceller with $L \in \{2, 3, \dots, 10\}$ and $P \in \{3, 5, 7, 9\}$. As shown in Fig.~\ref{fig:heatmap_poly_perf}, there is only a marginal difference in performance between the different polynomial cancellers, whereas the complexity quickly grows, as shown in Fig.~\ref{fig:heatmap_poly_complexity}. The maximum achievable SIC is $31.3$~dB and the lowest complexity model that comes within $1$~dB of this maximum uses $L=3$ and $P=7$ and achieves a SIC of $30.5$~dB (shown in Fig.~\ref{fig:heatmap_poly} with a circle).
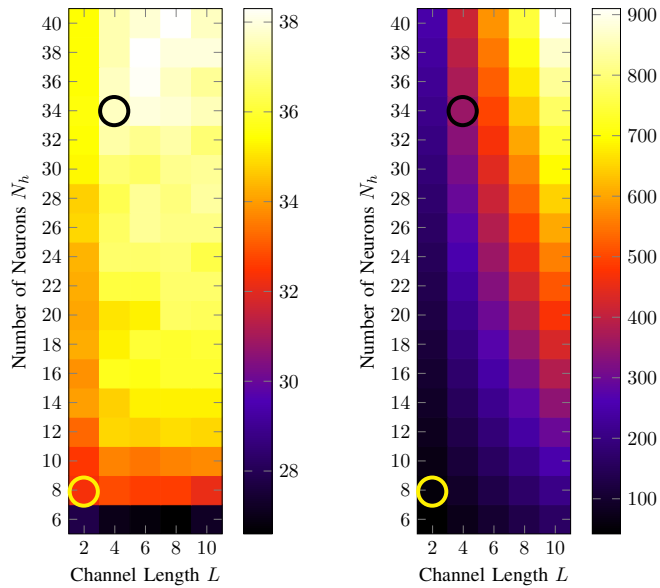
\begin{figure}[t]
	\centering
	\begin{subfigure}[t]{0.475\linewidth}
		\centering
		\scalebox{0.75}{\begin{tikzpicture}

\begin{axis}[
	width=4.3cm,
	height = 10.9cm,
    ylabel=Number of Neurons $N_h$,
    xlabel=Channel Length $L$,
    xtick={2,4,6,8,10,12},
    ytick distance=2,
    colormap/thermal,
%    colormap name=mymap3,
    colorbar,
    disabledatascaling, 
    enlargelimits=false,
    axis on top,
    ]
    % NB: The star inverts
   \addplot[matrix plot*,point meta=explicit] file [meta=index 2] {fig/data/heatmap_nn_perf.dat};

% For showing numbers   
%    \tiny
    % NB: The star inverts
%   \addplot[matrix plot*,point meta=explicit, nodes near coords] file [meta=index 2] {fig/data/heatmap_poly_perf.dat};   
\end{axis}

\draw[line width=2, yellow] (0.265, 0.75) circle (0.25); % Similar NN
\draw[line width=2] (0.805, 7.5) circle (0.25); % Peak NN
\end{tikzpicture}%
}
		\caption{SIC performance in dB.}
		\label{fig:heatmap_nn_perf}
	\end{subfigure}\hfill
	\begin{subfigure}[t]{0.475\linewidth}
		\centering
		\scalebox{0.75}{\begin{tikzpicture}

\begin{axis}[
	width=4.3cm,
	height = 10.9cm,
    ylabel=Number of Neurons $N_h$,
    xlabel=Channel Length $L$,
    xtick={2,4,6,8,10,12},
    ytick distance=2,    
    colormap/thermal,
%    colormap name=mymap3,
    colorbar,
    disabledatascaling, 
    enlargelimits=false,
    axis on top,
    ]
   \addplot[matrix plot*,point meta=explicit] file [meta=index 2] {fig/data/heatmap_nn_complexity.dat};
\end{axis}

\draw[line width=2, yellow] (0.265, 0.75) circle (0.25); % Similar NN
\draw[line width=2] (0.805, 7.5) circle (0.25); % Peak NN
\end{tikzpicture}%
}
		\caption{Number of multiplications.}
		\label{fig:heatmap_nn_complexity}
	\end{subfigure}
\caption{SIC performance and number of multiplications for the NN canceller as a function of the channel length $L$ and the number of neurons $N_h$ with $N_l=1$.  
The equi-performance NN canceller with $L=2$ and $N_h=8$ and the peak-performance NN canceller with $L=4$ and $N_h=34$ are marked with yellow and black circles, respectively.}
\label{fig:heatmap_nn}
\end{figure}
\begin{table}[t]
  \centering
  \caption{Comparison of the SIC performance and the complexity of the selected polynomial and NN-based cancellers. }\label{tab:perf}
  \begin{tabular}{lrrr}
  \toprule
                          & \multicolumn{1}{c}{Polynomial}  & \multicolumn{1}{c}{Equi NN}      & \multicolumn{1}{c}{Peak NN} \\
    \midrule
    Cancellation (dB)     & $30.5$ 	& $32.3$  & $37.6$ \\
    \midrule
    $L$                   & $3$  	& $2$    & $4$ \\
    $P$                   & $7$  	& n/a     & n/a \\
	$N_l$				  & n/a     & $1$    & $1$ \\
    $N_h$                 & n/a 	& $8$    & $34$ \\
    \midrule
	Real Add.        	  & $418$	& $82$   & {$428$} \\
    Real Mult.  		  & $180$ 	& $60$   & {$364$} \\
    \bottomrule
  \end{tabular}
\end{table}
In Fig.~\ref{fig:heatmap_nn}, we show two heatmaps for the SIC performance and the number of multiplications for NN cancellers with $L \in \{2, 4, \dots, 10\}$,  $N_h \in \{6, 8, \dots, 40\}$, and $N_l=1$, which were trained for $50$ epochs. We note that we also explored several architectures of deeper NNs (i.e., $N_l > 1$), but the shallow NN always achieved the same SIC performance with lower complexity.\footnote{Interestingly, in~\cite{Kristensen2019} it was shown that using a deep NN can be beneficial, but this result was obtained for a different dataset. This shows that a careful selection of the NN architecture based on the expected operating scenario is a useful tool to reduce the complexity.} 
%In order to reduce variation due to NN training, each value in the heatmaps is the result of picking the best model from three different random initializations.
The equi-performance NN has $L=2$, $N_h=8$, and achieves a SIC of $32.3$~dB (shown in Fig.~\ref{fig:heatmap_nn} with yellow circles). The peak-performance NN has $L=4$, $N_h=34$, and a SIC performance $37.6$~dB (shown in Fig.~\ref{fig:heatmap_nn} with black circles).
We summarize the above selection in Table~\ref{tab:perf}, where we also show the complexity of the SI cancellers in terms of the number of real-valued multiplications and additions given by \eqref{eq:addpoly}-\eqref{eq:multpoly} and \eqref{eq:addnn}-\eqref{eq:multnn}.
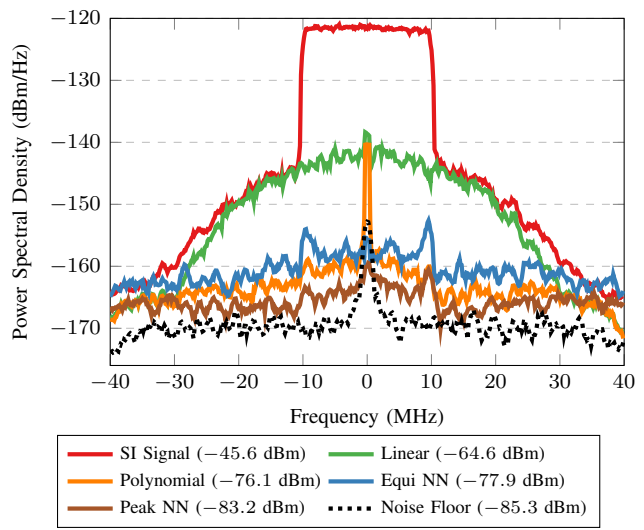
\begin{figure}[t]
  \centering
  \begin{tikzpicture}

	\pgfplotsset{grid style={dashed}}
	
	\begin{axis}[
		width = 0.95\columnwidth,
		height = 0.7\columnwidth,
		xlabel = {Frequency (MHz)},
		ylabel = {Power Spectral Density (dBm/Hz)},
		ylabel near ticks,
		xlabel near ticks,
		xtick distance=10,
		ytick distance=10,
		label style={font=\footnotesize},
		tick label style={font=\footnotesize},
		xmin = -40, xmax = 40,
		ymin = -176, ymax = -120,
		ymajorgrids,
		legend style={at={(0.425,-0.2)},anchor=north,font=\scriptsize},
		legend columns=2,
		legend cell align=left,
		legend entries={SI Signal ($-45.6$ dBm), Linear ($-64.6$ dBm), Polynomial ($-76.1$ dBm), Equi NN ($-77.9$ dBm), Peak NN ($-83.2$ dBm), Noise Floor ($-85.3$ dBm)}
	]		

		\addplot[Set1-7-1, ultra thick, solid, each nth point=8, filter discard warning=false] table[x index=0, y index = 1] {fig/data/NLyTestFFT.dat};
		\label{psd:nosic}
		\addplot[Set1-7-3, ultra thick, solid, each nth point=8, filter discard warning=false] table[x index=0, y index = 1] {fig/data/NLyTestLinCancFFT.dat};
		\addplot[Set1-7-5, ultra thick, solid, each nth point=8, filter discard warning=false] table[x index=0, y index = 1] {fig/data/NLyTestNonLinCancFFT.dat};
		\addplot[Set1-7-2, ultra thick, solid, each nth point=8, filter discard warning=false] table[x index=0, y index = 1] {fig/data/nn_ch2_nh8_yTestNonLinCancFFT.dat};
		\addplot[Set1-7-6, ultra thick, solid, each nth point=8, filter discard warning=false] table[x index=0, y index = 1] {fig/data/nn_ch4_nh34_yTestNonLinCancFFT.dat};
		\addplot[black, ultra thick, dotted, each nth point=8, filter discard warning=false] table[x index=0, y index = 1] {fig/data/NLnoiseFFT.dat};		

	\end{axis}

\end{tikzpicture}%
  \caption{Comparison of the SIC performance of polynomial and NN-based cancellers. The pre-digital-cancellation SI signal and the receiver thermal noise floor are also shown for comparison. Active RF cancellation is emulated using a $50$~dB attenuator at the receiver.}\label{fig:performance}
\end{figure}
\subsection{Self-Interference Cancellation Performance Comparison}
In Fig.~\ref{fig:performance}, we show the power spectral density (PSD) of the received SI signal $y_{\text{SI}}[n]$ before any SIC is performed, the PSD of the received signal when no transmission takes place (i.e., the \emph{effective} noise floor of the receiver), as well as the PSDs of the SI signals after linear SIC and non-linear SIC with the polynomial and NN-based cancellers shown in Table~\ref{tab:perf}. We observe that using the polynomial canceller or the equi-performance NN canceller results in a residual SI signal that is approximately $9$~dB above the receiver noise floor. While both cancellers achieve the same SIC performance, the PSDs of the residual SI signals are significantly different. In particular, the polynomial canceller does not model and, hence, can not cancel the carrier leakage around the DC tone, but it achieves a better SIC performance for the remaining in-band signal than the equi-performance NN canceller. The peak-performance NN canceller, on the other hand, can cancel the SI down to approximately $2.5$~dB from the receiver noise floor. This clearly shows that there are non-linear effects that cannot be modeled adequately by the polynomial canceller.
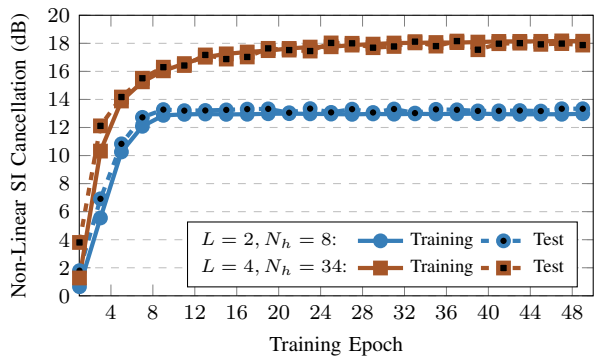
\begin{figure}[t]
  \centering
  \begin{tikzpicture}

	\pgfplotsset{grid style={dashed}}
	
	\begin{axis}[
		width = 0.95\columnwidth,
		height = 0.6\columnwidth,
		xlabel = {Training Epoch},
		ylabel = {Non-Linear SI Cancellation (dB)},
		ylabel near ticks,
		xlabel near ticks,
		xtick distance=4,
		ytick distance=2,
		label style={font=\footnotesize},
		tick label style={font=\footnotesize},
		xmin = 0.98, xmax = 50,
		ymin = 0, ymax = 20,
		ymajorgrids,
		legend pos = south east,
		legend style={font=\scriptsize},
		legend columns=3,
		legend cell align=left,
	]		

		% Shallow NN
		\addlegendimage{empty legend}
		\addlegendentry{$L=2,N_h=8$:\,}
		\addplot[Set1-7-2, ultra thick, solid, mark=*] table[x index = 0, y index = 1, each nth point=2] {fig/data/convergence_nn_ch2_nh8.dat};
		\addlegendentry{Training}
		\addplot[Set1-7-2, ultra thick, dashed, mark=*, mark options={solid,fill=black}] table[x index = 0, y index = 2, each nth point=2] {fig/data/convergence_nn_ch2_nh8.dat};
		\addlegendentry{Test}

		% Deep NN
		\addlegendimage{empty legend}
		\addlegendentry{$L=4,N_h=34$:\,}
		\addplot[Set1-7-6, ultra thick, solid, mark=square*] table[x index = 0, y index = 1, each nth point=2] {fig/data/convergence_nn_ch4_nh34.dat};
		\addlegendentry{Training}
		\addplot[Set1-7-6, ultra thick, dashed, mark=square*, mark options={solid,fill=black}] table[x index = 0, y index = 2, each nth point=2] {fig/data/convergence_nn_ch4_nh34.dat};
		\addlegendentry{Test}

	\end{axis}

\end{tikzpicture}%
  \caption{Training convergence of the equi-performance NN canceller with $L{=}2$ and $N_h{=}8$, as well as peak-performance NN canceller with $L{=}4$ and $N_h{=}34$.}\label{fig:convergence}
\end{figure}

In Fig.~\ref{fig:convergence}, we show the training convergence behavior for the non-linear cancellation part of the two NN cancellers.
The linear cancellation is in both cases approximately $19$~dB, making the non-linear SIC directly comparable. 
We observe that the equi-performance NN achieves its maximum performance on the test set after $9$ epochs, while the peak-performance NN requires more than $20$ epochs to achieve its maximum performance on the test set.
Moreover, we observe that both NN cancellers have similar performance on the training and test sets, meaning that there are no obvious overfitting issues. 
We note that in this work, we focus on the complexity of the inference part for both the polynomial canceller and the NN-based cancellers. However, the complexity of the training part is an important issue that should also be carefully considered, even though training is typically required much less often than inference.
\section{Hardware Implementation Results}\label{sec:hardwareresults}
In this section, we present a comparison of FPGA and ASIC implementation results for the polynomial SI canceller and the NN-based SI cancellers.
\subsection{Comparison Setup}
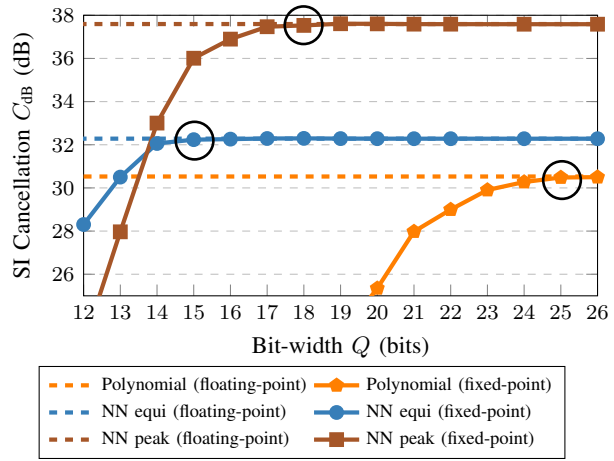
\begin{figure}[t]
  \centering
  \begin{tikzpicture}

	\pgfplotsset{grid style={dashed}}

	\begin{axis}[
		width = 0.95\columnwidth,
		height = 0.6\columnwidth,
		xlabel = {Bit-width $Q$ (bits)},
		ylabel = {SI Cancellation $C_{\text{dB}}$ (dB)},
		ylabel near ticks,
		xlabel near ticks,
		xtick distance=1,
		ytick distance=2,
		label style={font=\small},
		tick label style={font=\footnotesize},
		xmin = 12, xmax = 26,
		ymin = 25, ymax = 38,
		ymajorgrids,
		legend style={at={(0.425,-0.25)},anchor=north,font=\scriptsize},
		legend columns=2,
		legend cell align=left,		
		legend entries={Polynomial (floating-point),
                    Polynomial (fixed-point),
					NN equi (floating-point),
                    NN equi (fixed-point),
					NN peak (floating-point),
                    NN peak (fixed-point),
}                    
	]

		\addplot[Set1-7-5, ultra thick, dashed] table[x index = 2, y index = 3] {fig/data/bitwidth_power_poly.dat};
		\addplot[Set1-7-5, ultra thick, solid, mark=pentagon*, mark options={solid, ultra thick}] table[x index = 0, y index = 1] {fig/data/bitwidth_power_poly.dat};

		\addplot[Set1-7-2, ultra thick, dashed] table[x index = 2, y index = 3] {fig/data/bitwidth_power_nn_similar.dat};
		\addplot[Set1-7-2, ultra thick, solid, mark=*, mark options={solid, ultra thick}] table[x index = 0, y index = 1] {fig/data/bitwidth_power_nn_similar.dat};

		\addplot[Set1-7-6, ultra thick, dashed] table[x index = 2, y index = 3] {fig/data/bitwidth_power_nn_peak.dat};
		\addplot[Set1-7-6, ultra thick, solid, mark=square*, mark options={solid, ultra thick}] table[x index = 0, y index = 1] {fig/data/bitwidth_power_nn_peak.dat};

	\end{axis}

	\draw[line width=1] (1.475,2.065) circle (0.25); % Small NN
	\draw[line width=1] (6.36,1.54) circle (0.25); % Poly
%	\draw[line width=1] (3.1,4.02) circle (0.25);	
	\draw[line width=1] (2.924,3.6) circle (0.25); % Big NN
\end{tikzpicture}%
  \caption{Total SIC for the polynomial and NN SI cancellers as a function of the datapath bit-width $Q$. The circled points for each canceller are used for the FPGA and ASIC implementation results in Section~\ref{sec:hardwareresults}.}\label{fig:quantization}
\end{figure}

To perform a meaningful comparison of FPGA and ASIC implementation results, the quantization bit-width $Q$ for the different cancellers needs to be selected to individually minimize the implementation complexity while keeping the performance of the SI cancellers as close as possible to their floating-point equivalents. 
In Fig.~\ref{fig:quantization}, we show the cancellation performance for the polynomial SI canceller and the NN SI cancellers as a function of the quantization bit-width $Q$. 
We observe that both NN SI cancellers generally require a lower quantization bit-width $Q$ compared to the polynomial SI canceller to achieve SIC performance comparable to the floating-point performance.

Moreover, for the hardware implementation results presented in this section, we choose $Q=16$ for the equi-performance NN canceller, $Q=18$ for the peak-performance  NN canceller, and $Q=25$ for the polynomial SI canceller, as this choice leads to effectively identical SIC performance as the corresponding floating-point implementations for all cancellers. 
We note that the peak performance NN SI canceller requires one additional integer bit compared to the equi-performance NN SI canceller, due to larger absolute output values in the hidden layer.
For the equi-performance NN canceller, we set $N_{\text{PE},1} = 8$ and $N_{\text{PE},2} = 4$ so that $\mathcal{T}_1 = \mathcal{T}_2 = \nicefrac{1}{4}$. 
With this setting, the macro-pipeline is perfectly balanced and one SI cancellation sample is produced every $4$ clock cycles. 
Furthermore, $N_{\text{CPE},\text{linear}} = 1$ CPEs are instantiated for the NN SI canceller to ensure that the linear cancellation step can be completed in the same number of cycles. 
For the peak-performance NN canceller, we set $N_{\text{PE}} = 40$ for the hidden layer and $N_{\text{PE}} = 10$ for the output layer so that the throughput for both layers is $\mathcal{T} = \nicefrac{1}{7}$. 
Again, we use $N_{\text{CPE},\text{linear}} = 1$ for the linear canceller.
The equi-performance NN canceller thus requires a total of $12$ PEs and the peak-performance NN canceller requires a total $50$ PEs, and both require only $1$ CPE for the linear canceller.
Finally, for the polynomial canceller, we use $N_{\text{CPE}} = 10$ complex PEs. We use relatively high parallelization because our cancellers need to achieve a throughput at least equal to the $80$~Msamples/s sampling frequency that is used in our dataset.

\subsection{FPGA Implementation Results}\label{sec:resultsfpga}
In Table~\ref{tab:fpga-res}, we show place-and-route (PAR) results on a Xilinx Virtex-7 XC7VX485 (speed grade -2) FPGA, which contains a total of $75.9$k slices, $303.6$k LUTs, $607.2$k flip-flops, and $2.8$k DSP slices. A clock frequency target of $100$~MHz is used for all cancellers.

We observe that the equi-performance NN canceller has the smallest resource utilization of all considered cancellers, while the peak-performance NN canceller has a similar resource utilization to the polynomial canceller, while providing approximately $7$~dB better SIC performance. Moreover, the equi-performance NN canceller and the peak-performance NN canceller have a $95$\% and $5$\% higher throughput than the polynomial SI canceller. However, when implemented on an FPGA, none of the considered SI cancellers can achieve the $80$~Msamples/s throughput that is required by the considered application.
\begin{table}[t]
  \setlength{\tabcolsep}{5pt}
  \centering
  \caption{FPGA implementation Results (Virtex-7 XC7VX485).}\label{tab:fpga-res}
  \begin{tabular}{lrrr}
  \toprule
							& \multicolumn{1}{c}{Poly.} 	& \multicolumn{1}{c}{Equi NN}	& \multicolumn{1}{c}{Peak NN} \\
    \midrule
    Slices            		& $2244$ ($2.96$\%)  		& $\mathbf{514}$ ($0.68$\%)  		& $1619$ ($2.13$\%) \\
    ~~~LUT (logic)    		& $5422$ ($1.79$\%)  		& $\mathbf{793}$ ($0.26$\%) 		& $2462$ ($0.81$\%) \\
    ~~~LUT (RAM)      		& ${946}$ ($0.31$\%)  		& $\mathbf{336}$ ($0.11$\%) 			& $1506$ ($0.50$\%) \\
    ~~~Registers      		& ${2320}$ ($0.38$\%)   	& $\mathbf{887}$ ($0.15$\%)  		& $2142$ ($0.35$\%) \\
    ~~~DSP Slices     		& ${84}$  ($3.00$\%)   		& $\mathbf{15}$ ($0.54$\%)    		& ${53}$ ($1.89$\%) \\
	\midrule
    Frequency (MHz)      	& $87.11$     				& $\mathbf{96.86}$         		& $91.84$  \\
    T/P (Msamples/s) 		& $12.44$      				& $\mathbf{24.22}$             		& ${13.12}$  \\
	T/P (samples/cycle)		& $\nicefrac{1}{7}$			& $\mathbf{\nicefrac{\mathbf{1}}{\mathbf{4}}}$ & ${\nicefrac{1}{7}}$ \\
	Latency (ns)	  		& $91.84$					& $\mathbf{51.62}$			  		& $87.11$ \\
	Latency (cycles)  		& $8$						& $\mathbf{5}$	 					& $8$ \\
    \bottomrule
  \end{tabular}
\end{table}

\subsection{ASIC Implementation Results}\label{sec:resultsasic}
In Table~\ref{tab:asic}, we present ASIC implementation results for the polynomial SI canceller and the two NN SI cancellers using a 28~nm FD-SOI technology. We use typical-typical corners, a $0.9$~V operating voltage, and a $25$\degree~C operating temperature. The polynomial canceller and the peak-performance canceller were synthesized, placed, and routed for a target frequency of $400$~MHz and $1$~GHz, respectively. However, for the power results, all cancellers are operated at a frequency that results in a throughput of exactly $80$~Msamples/s, i.e., $560$~MHz for the polynomial canceller and the peak-performance NN canceller, and $320$~MHz for the equi-performance NN canceller. Moreover, post-PAR simulations are used both to verify the design and to accurately estimate the switching activity.

We observe that the equi-performance NN canceller requires a significant $8.1\times$ less area and  $7.7\times$ less power than the polynomial canceller. We note that the absolute latency of the equi-performance NN canceller is $0.9$~ns higher than the polynomial canceller, but this difference is negligible in practice. The peak-performance NN canceller, on the other hand, is $1.2\times$ smaller than the polynomial canceller and requires slightly more ($1.3\times$) power. However, it should be noted that the peak-performance NN canceller also has an approximately $7$~dB better SIC performance than the polynomial canceller.
\section{Conclusion}\label{sec:conclusion}
In this paper, we presented a high-throughput hardware architecture for a NN-based SIC scheme for full-duplex radios. We also presented, to the best of our knowledge, the first efficient hardware architecture for polynomial SIC in the literature, which we used as a comparison baseline for the NN-based SI cancellers. Our implementation results show that the NN SI cancellers have significantly lower computational complexity than a conventional polynomial SI canceller, which translates into substantial area and energy savings when the schemes are implemented in hardware. Specifically, for the same SIC performance, an ASIC implementation of a NN-based SI canceller has up to $8.1\times$ and $7.7$ better hardware efficiency and energy efficiency when compared to a conventional polynomial SI canceller.
\begin{table}[t]
	\centering
	\caption{ASIC Implementation Results ($28$~nm FD-SOI, typical-typical corners, $0.9$~V, $25$\degree~C).}\label{tab:asic}
	\begin{tabular}{lrrr}
	\toprule
		~ 										& Poly.				& Equi NN 			& Peak NN 	\\
		\midrule
		Area (mm$^2$)         					& ${0.179}$  		& $\mathbf{0.022}$   		& ${0.150}$ \\
		Area (kGE)         						& ${364.6}$ 		& $\mathbf{44.4}$ 			& ${306.7}$	 \\	
		\midrule   		
		Frequency (MHz)         				& ${560}$   		& $320$   			& ${560}$ 	 \\
		Throughput (Msamples/s)       			& $80$ 				& $80$    			& ${80}$ 		\\
		Throughput (samples/cycle)				& $\nicefrac{1}{7}$	& $\mathbf{\nicefrac{\mathbf{1}}{\mathbf{4}}}$	& ${\nicefrac{1}{7}}$ \\
		Latency (ns)							& $\mathbf{14.3}$			& ${15.6}$			& $\mathbf{14.3}$ 		\\
		Latency (cycles)						& $8$				& $\mathbf{5}$				& $8$ 		\\
		\midrule    
		Total Power (mW)      					& ${84.14}$ 		& $\mathbf{10.95}$   		& ${112.70}$ \\            
		~~~Internal Power (mW)    				& ${42.78}$  		& $\mathbf{6.31}$   			& ${64.78}$ \\
		~~~Switching Power (mW)   				& ${41.36}$  		& $\mathbf{4.63}$    		& ${47.80}$ 		 \\
		~~~Leakage Power (mW)    				& $0.06$   			& $\mathbf{0.01}$    		& ${0.07}$ 	 \\
		\midrule
		Hardware Eff. (Msamples/s/mm$^2$)  	& $448$ 		& $\mathbf{3\,679}$   	& ${533}$  \\
		Energy Eff. (nJ/sample)   			& $1.05$    	& $\mathbf{0.14}$  		& ${1.41}$  \\
		\bottomrule
	\end{tabular}
\end{table}
%\bstctlcite{IEEEexample:BSTcontrol}
%\bibliographystyle{IEEEtran}
%\bibliography{IEEEabrv,bibliography}
%

\vspace{0.5cm}
% Bios
\begin{IEEEbiography}[{\includegraphics[width=1in,height=1.25in,clip,keepaspectratio]{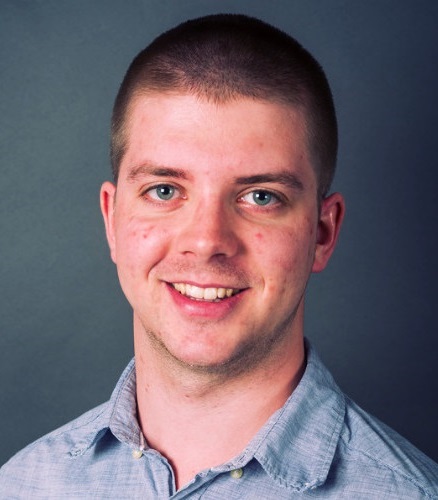}}]%
{Yann Kurzo} received the B.Sc. degree from the Haute \'ecole d'ing\'enierie et d'architecture (HEIA-FR) in 2014 and the M.Sc. degree in Electrical Engineering from the Ecole polytechnique f\'ed\'erale de Lausanne (EPFL) in 2018. He is currently a Digital Design Engineer at ON Semiconductor in Marin, Switzerland. 
\end{IEEEbiography}
\vspace{-1cm}
\begin{IEEEbiography}[{\includegraphics[width=1in,height=1.25in,clip,keepaspectratio]{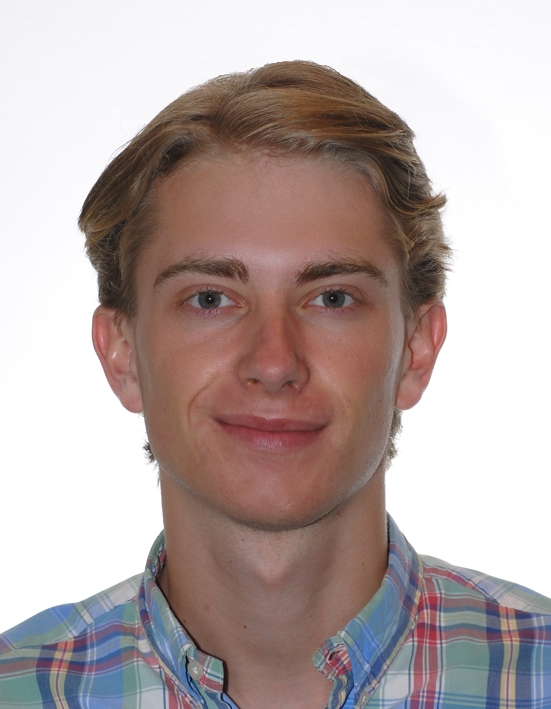}}]%
{Andreas Toftegaard Kristensen}
 (Student Member, IEEE) was born in Hiller\o{}d, Denmark, in 1994. 
He received his B.Sc. degree in Electrical Engineering in 2017 and his M.Sc. degree (Honours) in Computer Science and Engineering in 2019, both from the Technical University of Denmark.
He is currently pursuing a Ph.D. degree in Electrical Engineering under the supervision of Prof. Andreas  Burg in the Telecommunications Circuits Laboratory at EPFL, Switzerland. 
His research interests include vital-sign detection using commodity WiFi routers, full-duplex self-interference cancellation, and custom hardware architectures for neural networks.
\end{IEEEbiography}
\vspace{-1cm}
\begin{IEEEbiography}[{\includegraphics[width=1in,height=1.25in,clip,keepaspectratio]{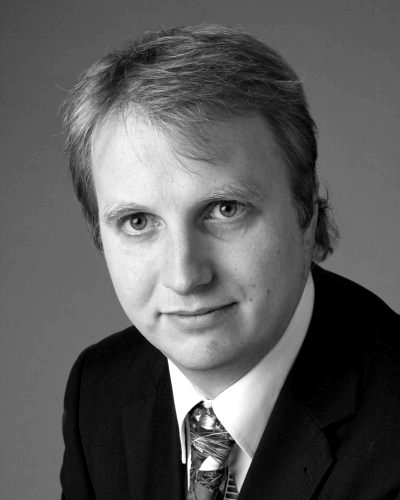}}]%
{Andreas Burg} (Member, IEEE) was born in Munich, Germany, in 1975. He received the Dipl.-Ing. degree from the Swiss Federal Institute of Technology (ETH) Zurich, Zurich, Switzerland, in 2000, and the Dr. sc. techn. degree from the Integrated Systems Laboratory, ETH Zurich, in 2006.
In 1998, he worked at Siemens Semiconductors, San Jose, CA, USA. During his doctoral studies, he worked at Bell Labs Wireless Research for one year. From 2006 to 2007, he was a Post-Doctoral Researcher with the Integrated Systems Laboratory and with the Communication Theory Group, ETH Zurich. In 2007, he co-founded Celestrius, an ETH-spinoff in the field of MIMO wireless communication, where he was responsible for the ASIC development as the Director for VLSI. In January 2009, he joined ETH Zurich as a SNF Assistant Professor and as the Head of the Signal Processing Circuits and Systems Group, Integrated Systems Laboratory. In January 2011, he joined the Ecole polytechnique f\'ed\'erale de Lausanne (EPFL), where he is leading the Telecommunications Circuits Laboratory. He was promoted to Associate Professor with tenure in June 2018.
Dr. Burg is a member of the EURASIP SAT SPCN, the IEEE TC-DISPS, and the CAS-VSATC. He has served on the TPC of various conferences on signal processing, communications, and VLSI. He was a TPC Co-Chair for VLSI-SoC 2012 and ESSCIRC 2016 and SiPS 2017. He was a General Chair of ISLPED 2019. He served as an Editor for the IEEE Transaction of Circuits and Systems in 2013 and on the Editorial Board of the Springer Microelectronics Journal. He is currently an Editor of the Springer Journal on Signal Processing Systems, MDPI Journal on Low Power Electronics and Applications, and the IEEE Transactions on Very Large Scale Integration (VLSI) Systems.
\end{IEEEbiography}
\vspace{-1cm}
\begin{IEEEbiography}[{\includegraphics[width=1in,height=1.25in,clip,keepaspectratio]{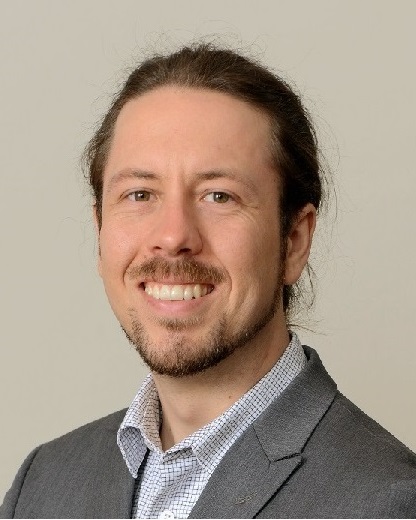}}]%
{Alexios Balatsoukas-Stimming} (Member, IEEE) received the Diploma and M.Sc. degrees in electronics and computer engineering from the Technical University of Crete, Chania, Greece, in 2010 and 2012, respectively, and the Ph.D. degree in computer and communications sciences from the Ecole polytechnique f\'ed\'erale de Lausanne (EPFL), Switzerland, in 2016. He then spent one year at the European Laboratory for Particle Physics (CERN) as a Marie Skłodowska-Curie Post-Doctoral Fellow. He was a Post-Doctoral Researcher with the Telecommunications Circuits Laboratory, EPFL, from 2018 to 2019. He is currently an Assistant Professor with the Eindhoven University of Technology, The Netherlands. His research interests include VLSI circuits for signal processing and communications, error correction coding theory and practice, as well applications of machine learning to signal processing for communications. 
\end{IEEEbiography}

\end{document}